# Feedback Microrheology in Soft Matter

Kenji Nishizawa,‡[b] Natsuki Honda,†[a] Masahiro Ikenaga,[a] Yujiro Sugino,[a] Shono Inokuchi,[a] Hiroyuki Ebata,[a] Takayuki Ariga[c] and Daisuke Mizuno*[a]

Soft matter consists of meso-scale (nm~μm) structures that are formed by weak interactions and reorganized with thermal activations. The relaxation processes that occur spontaneously in such materials may be probed with microrheology, by observing the movement of embedded probe particles. Because of the softness of the material, however, perturbations to the probe that are inevitably added during microrheology experiments prevent direct translation of those movements to rheological properties. In this study, we conducted optical-trap-based microrheology with significantly reduced mechanical perturbations. With dual feedback technology, well-determined optical-trapping forces were applied to a fluctuating embedded probe, and its response and fluctuations were precisely measured with high spatiotemporal resolution. We demonstrate the improved performance of this technique in an reconstituted network of actin cytoskeletal filaments, by observing their slow dynamics, homogeneous thermal fluctuations as well as activated hopping between mesoscale microenvironments. We discuss how heterogeneous relaxations observed in equilibrium soft matter become homogeneous under constant forcing beyond linear regime.

## Introduction

Rheological properties are some of the most fundamental physical attributes of a material. Over many years, they have been extensively investigated via experiments, computer simulations and analytic theories. Nowadays, the purely elastic or viscous response of "simple" materials is understood at least qualitatively based on the microscopic processes that may plausibly be occurring in the system [1-3]. However, in nature, the rheological behavior of "soft" materials is more complex and typically nonlinear. For example, polymer networks [4,5] and colloidal suspensions [6,7] show a variety of relaxation spectra in their linear response. Under moderate mechanical loads, soft matter can either weaken (fluidize) [8,9], stiffen [10,11], or experience both in complex ways [12,13] depending on the details of the materials and experimental protocols. Understanding this complex behavior has long been the focus of studies in soft matter physics.

Mechanical properties of soft materials are associated with mesoscale (nm~μm) internal structures, such as the mesh of polymer networks, the persistence of semi-flexible filaments, and the excluded volume of colloidal components. Weak interactions (*e.g.* van der Waals, electrostatic, steric, and hydrodynamic) between these mesoscale structural units define the softness of their macroscopic mechanical response. Owing to this softness, gentle thermal/external forcing induces vigorous fluctuations that relax over timescales that typically extend from less than μs to more than days. Thus, the mechanical properties of soft matter intriguingly depend on both time and length scales [14-20]. In order to better understand the complex mechanics of soft matter, it is therefore crucial to apply well-controlled forces to mesoscale structures of interest and observe their response with high spatiotemporal resolution[18,19]. Recent progress in micro- and nano-technology tools have opened the door to perform such rheology experiments; these are collectively called microrheology (MR).

MR is a technique to probe the local mechanical properties of a sample from the movement of embedded probe particles [21-26]. Linear MR observes either a probe's spontaneous fluctuations (Passive MR: PMR)[5,22,26] or its response to small external forces (Active MR: AMR)[23,27]. Since probe dynamics depend on the viscoelastic resilience of the surrounding material, MR translates probe movements to the viscoelastic shear modulus of the surrounding medium. At thermodynamic equilibrium, the fluctuation-dissipation theorem (FDT) verifies that AMR and PMR provide equivalent information *i.e.* the linear viscoelasticity of surrounding media

[a.] *Department of Physics, Kyushu University, 819-0395 Fukuoka, Japan. E-mail: mizuno@phys.kyushu-u.ac.jp*
[b.] *Institute of Developmental Biology of Marseille, Campus de Luminy case 907, 13288 Marseille Cedex 09, France.*
[c.] *Graduate School of Medicine, Yamaguchi University, Ube, Japan*

[23, 28, 29]. Conversely, forcing the probe particle *beyond* the linear regime drives the system artificially out of equilibrium and allows investigation of the nonlinear response of soft matter. Such nonlinear MR can be conducted, for instance, by applying a strong constant force to the probe particle [30, 31].

MR experiments have been performed with high bandwidth and high precision by utilizing optical traps and laser interferometry techniques [21, 22, 26]. The probe position is measured via the diffraction of a weak probe laser that impinges on the particle (the back-focal-plane interferometry: BFPI [32]). For AMR, a small sinusoidal force is applied to the tracer particle by another optical trap provided by a separate driving laser [23-25]. In conventional linear MR experiments, however, optical-trapping forces are not well controlled. Since trapping forces were imposed via an open-loop operation, they fluctuated together with the random fluctuation of the colloidal particles themselves. Furthermore, the probe movements were suppressed by the trapping potential $U(r) = 1/2 kr^2$ formed around the laser focus, where $k$ is the trap stiffness, and $r$ is the distance from the laser focus. When probe movements are suppressed owing to the artificial potential, the experimental bandwidth is typically limited at low frequencies to ~ 1 Hz. Exerting not-well-controlled forces to a probe particle is inappropriate especially for nonlinear MR since it may induce complex non-linear and non-equilibrium dynamics in a way that prevents theoretical analysis.

The first half of this article focuses on the technical progress which circumvents these problems for linear MR. We introduced a fast feedback to control the position of the drive laser so that it quickly follows the fluctuating probe particle. The position (focus) of the drive laser was rapidly optimized by feedback control of the acousto-optic deflector (AOD) using measurements of the probe particle's position taken via laser interferometry. This feedback technique, referred to as force feedback, allows us to apply a well-controlled force to the fluctuating probe. We investigated the performance of the force feedback for linear MR by observing thermal fluctuations of a probe particle dispersed in simple liquids with well-determined viscosity. With and without force feedback, the position of the feedback-controlled laser, probe displacements from the laser focus, and the total displacements were analyzed in detail. It was verified that the probe particle does not feel the optical potential if the feedback-response time is smaller than the characteristic time ($\gamma_0 / k$) of the probe fluctuation in the trapping potential [23]. Here, $\gamma_0$ is the friction coefficient of the probe. We then conducted force-feedback AMR in soft matter under thermal equilibrium (entangled F-actin hydrogel). By comparing the results under force feedback with those found via conventional MR without feedback, we demonstrate that the force feedback provides more reliable and accurate data at low frequencies even if a strong drive laser is used.

In the second half of this article, we describe nonlinear MR experiments which were conducted by applying constant forces to the probe particle, utilizing the force-feedback technique. In this way, the soft material surrounding the probe (here, a sparsely crosslinked actin gel) was stably forced beyond its linear response regime. Along with the local non-linear response, the structural relaxation was enhanced by the forcing the soft material, leading to large fluctuations and drift motion of the probe particle. In order to investigate these slow non-equilibrium relaxations, it was necessary to track the probe particle drifting over large distances. However, solely applying force feedback does not work for that purpose because the errors and uncertainties in the BFPI measurement increases when the feedback-controlled laser moves away from the optical axis. In this study, we therefore introduced another feedback referred to as *stage feedback* [33]. In stage feedback, the piezo-mechanical stage on which the sample chamber was placed was feedback controlled while the probe position was measured with the laser interferometry of the fixed probe laser. This technique keeps a probe particle around the probe laser focus even if the particle is largely drifting/fluctuating in a specimen. By operating these two feedback techniques simultaneously (dual feedback mode), we succeeded in applying a well-controlled force to a drifting probe particle while keeping the particle around the center of the optical axis. The potential of the developed technique was demonstrated in loosely crosslinked F-actin gels. Forces of up to several pN caused directed movements of the probe, not a continuous smooth movement like in homogeneous liquids but with intermittent jumps that occurred randomly both in time and size. Despite the apparent heterogeneity, a careful statistical analysis in fact showed that the underlying energy landscape was *homogeneous*ly stochastic. These findings clearly highlight the potential of the developed technique to investigate nonlinear and nonequilibrium dynamic responses in soft matter.

# 1. Linear Feedback Microrheology

## 1.1 Conventional MR

In conventional MR, the shear viscoelastic modulus. $G(\omega)$. of a soft material is obtained by measuring the displacement response of an embedded probe particle $u(t)$ to an externally-applied force $F(t)$ and a thermally fluctuating force $f(t)$. The probe movements in an equilibrated specimen can be described by the Langevin equation,

$$\int_{-\infty}^{t} \gamma(t-t')\dot{u}(t')dt' = F(t) + f(t). \qquad (1)$$

Here $\dot{u}(t)$ is the velocity of the probe, and $\gamma(t)$ is a friction function that reflects memory due to the viscoelastic resilience of surrounding material. In the case of linear AMR, a small sinusoidal force $F(t) = \hat{F}(\omega)e^{-i\omega t}$ is applied, and the displacement response $u(t) = \hat{u}(\omega)e^{-i\omega t}$ is measured. By taking the ensemble average of eqn (1) for a periodic steady-state, the frequency-dependent response function $\alpha(\omega)$ of the probe displacement is then given by

$$\alpha(\omega) \equiv \langle \hat{u}(\omega) \rangle / \hat{F}(\omega) = -1/[i\omega\tilde{\gamma}(\omega)]. \quad (2)$$

Here and hereafter, $\tilde{\ }$ and $\hat{\ }$ denote a Fourier-transformed function and the magnitude of a sinusoidal function which is synchronous to the applied external field, respectively. Note that $\hat{u}(\omega)$ has a complex quantity to represent the phase delay. The angled brackets denote a statistical or time average. For PMR, the probe's thermal fluctuation $u(t)$ is measured without an external force, i.e. $F(t) = 0$. Calculating the power spectral density ($\langle |\tilde{u}(\omega)|^2 \rangle \equiv \int \langle u(t)u(0) \rangle e^{i\omega t} dt$), the imaginary part of the response function $\alpha''(\omega)$ is obtained via the the fluctuation-dissipation theorem $\alpha''(\omega) = \omega\langle |\tilde{u}(\omega)|^2 \rangle / 2k_{\mathrm{B}}T$ (FDT). Since the real part $\alpha'(\omega)$ can be calculated from $\alpha''(\omega)$ by using Kramers-Kronig relation, $\alpha(\omega)$ is obtained as a full complex quantity. The complex shear modulus is obtained by the generalized Stokes relation as

$$G(\omega) = 1/6\pi a\alpha(\omega). \quad (3)$$

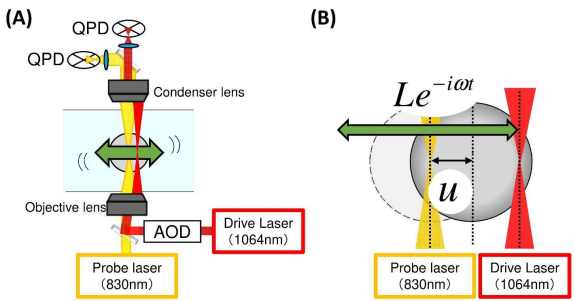

**Fig. 1** Conventional MR
(A) Schematic of the setup for conventional MR. An external force is applied by the AOD-controlled drive laser (red) and the displacement of the probe particle is detected by a probe laser (yellow). The deflection of each laser is detected by a QPD placed in the back focal plane of the condenser and objective lenses. (B) The laser and probe displacements in the focal plane. In AMR, the drive laser is sinusoidally oscillated around the focus of the fixed probe laser.

Whereas the theoretical basis of MR summarized as above looks simple, the actual execution of experiments requires more careful considerations. Conventional AMR and PMR [21-23] have been typically performed by applying an external force with the optical trap, and by measuring the probe displacements using a technique referred to as back-focal-plane laser interferometry (BFPI) [32]. In BFPI, a quadrant photodiode (QPD) placed at the back-focal plane of the objective and condenser lenses detects the interferometric pattern of the laser deflected by the trapped probe particle (Fig.1 A). For typical conventional AMR, the external force is applied by an AOD-controlled drive laser whose focus position is oscillated by $Le^{-i\omega t}$ from its average position (Fig.1 B). The displacenement response of the probe is measured with BFPI using another fixed probe laser. The Langevin equation for the probe particle in this situation is described as

$$k_{\mathrm{p}}u(t) + \int_{-\infty}^{t} \gamma(t-t')\dot{u}(t')dt' = k_{\mathrm{d}}\left(Le^{-i\omega t} - u(t)\right) + f(t), \quad (4)$$

where $k_{\mathrm{p}}$ and $k_{\mathrm{d}}$ are the trap stiffness of the probe and the drive laser, respectively, and $L$ is the amplitude of the drive laser oscillation. In conventional AMR, the optical-trapping force $F(t) = k_{\mathrm{d}}Le^{-i\omega t} - (k_{\mathrm{d}} + k_{\mathrm{p}})u(t)$ is not well controlled since $u(t)$ thermally fluctuates and its average frequency response $\langle \hat{u}(\omega) \rangle$ depends on the mechanical properties of the surrounding medium which are unknown prior to measurements. For the sake of convenience, we define the apparent driving force as $F_{\mathrm{apr}}(t) = \hat{F}_{\mathrm{apr}}(\omega)e^{-i\omega t} \equiv k_{\mathrm{d}}Le^{-i\omega t}$ and the apparent response function as

$$A(\omega) \equiv \langle \hat{u}(\omega) \rangle / \hat{F}_{\mathrm{apr}}(\omega) = \alpha(\omega)/\left[1 + (k_{\mathrm{p}} + k_{\mathrm{d}})\alpha(\omega)\right]. \quad (5)$$

Here, $\hat{F}_{\mathrm{apr}}(\omega) = k_{\mathrm{d}}L$ is the amplitude of the apparent driving force. The intrinsic response function $\alpha(\omega)$ is then obtained by solving eqn (5) [23]. However, when the optical-trapping is strong and/or the oscillation frequency $\omega$ is low, i.e. $k_{\mathrm{p}} + k_{\mathrm{d}} \gg 1/|\alpha(\omega)|$, eqn (5) hardly depends on $\alpha(\omega)$. In this case, the estimation of $\alpha(\omega)$ causes large errors. Note that the probe particle in this condition is mostly located at the bottom of the averaged potential created by the two lasers, $u(t) \sim k_{\mathrm{d}}Le^{-i\omega t}/(k_{\mathrm{d}} + k_{\mathrm{p}})$, i.e., the second term in the L.H.S. of eqn (4) is negligible. It is then hard to estimate the very small force amplitude $\langle \hat{F}(\omega) \rangle = k_{\mathrm{d}}L - (k_{\mathrm{d}} + k_{\mathrm{p}})\langle \hat{u}(\omega) \rangle \sim 0$ from the measured quantities: $k_{\mathrm{p}}, k_{\mathrm{d}}, L$, and $\langle \hat{u}(\omega) \rangle$. This problem determines the low frequency limit of conventional AMR experiments.

In conventional PMR, the drive laser is turned off and the spontaneous fluctuation of the probe particle is measured by the BFPI using a fixed probe laser. The apparent response function $A(\omega)$ is now obtained by substituting $k_{\mathrm{d}} = 0$ into eqn (5) as [23]

$$A(\omega) = \alpha(\omega)/\left[1 + k_{\mathrm{p}}\alpha(\omega)\right]. \quad (6)$$

The FDT under this condition is given as

$$\langle |\tilde{u}(\omega)|^2 \rangle^{\mathrm{trap}} = \frac{2k_{\mathrm{B}}TA''(\omega)}{\omega}, \quad (7)$$

where $\langle |\tilde{u}(\omega)|^2 \rangle^{\mathrm{trap}}$ is the power spectral density (PSD) of the trapped particle. After calculating the real part $A'(\omega)$ using the Kramers-Kronig relation, the apparent response function $A(\omega)$ is obtained. Equation (6) can then be solved to obtain the intrinsic response function $\alpha(\omega)$ [23]. However, with a strong trap and/or at low frequencies, *i.e.* $\alpha(\omega) = -1/[i\omega\tilde{\gamma}(\omega)] \gg 1/k_p$ , the R.H.S. of eqn (6) is approximately $1/k_p$ and seldom depends on $\alpha(\omega)$ . Therefore, $\alpha(\omega)$ estimated in this way exhibits large errors.

It is therefore necessary to use weak lasers in order to conduct the optical-trapping-based MR in soft materials. However, decreasing the laser power causes other problems. For instance, shot noise arises at high frequencies in BFPI signals, owing to the lack of photons for the QPD. Even at equilibrium, the probe can escape the weak trap due to Brownian motion during an experiment. Conventional MR contains fundamental limitations for investigating the slow dynamics of soft matter.

## 1.2 Force-feedback MR

### 1.2.1 Force-feedback PMR

In the case of PMR, the above-mentioned problems are circumvented if the optical-trapping force $F(t)$ is tuned to zero regardless of the stochastically fluctuating movement of the probe particle. In the case of PMR, this is achieved by introducing a fast feedback to the focus position of the probe laser. As shown in Fig. 2A, a $\lambda = 1064$ nm laser, which is now used as a probe laser, is controlled by an AOD so that it rapidly tracks the probe particle. PMR under such feedback control, which is referred to as the force-feedback PMR, was performed as shown below.

We define displacements $u(t)$ , $u_{AOD}(t)$ and $u_d(t)$ as shown in Fig. 2B. $u(t)$ and $u_{AOD}(t)$ are the displacements of the probe particle and the focus of the laser, respectively. $u_d(t)$ is the distance between the probe particle and the focus of the laser. The displacement of the probe in the sample is described as $u(t) = u_{AOD}(t) - u_d(t)$ . As shown in Fig. 2A, the output voltage $V(t)$ of the QPD, which is proportional to $u_d(t)$ , was fed to a Proportional-Integral-Derivative (PID) controller while the set point is grounded (Fig. 2A). An output signal $\varepsilon_{AOD}(t)$ from the PID controller was produced via the integral term of the PID, $\varepsilon_{AOD}(t) = I\int V(t) - s(t)\,dt$ , where $I$ is the programmable feedback gain and $s(t)$ is the set point of the PID controller. In force-feedback PMR, the set point is grounded, $s(t) = 0$ . The laser is deflected by the AOD by an amount proportional to $\varepsilon_{AOD}(t)$ . Therefore $u_{AOD}(t) \propto \varepsilon_{AOD}(t)$ holds as long as the response of the laser deflection to $\varepsilon_{AOD}(t)$ is sufficiently fast. $u_{AOD}(t)$ and $u_d(t)$ are then correlated via the feedback as,

$$u_{AOD}(t) = C_{AOD}\varepsilon_{AOD}(t) = (1/\tau_{PID})\int u_d dt. \quad (8)$$

The proportionality constants, $C_d = u_d(t)/V(t)$ and $C_{AOD} \equiv u_{AOD}(t)/\varepsilon_{AOD}(t)$ , were obtained following procedures given in Refs. [23, 33]. Here, $\tau_{PID} \equiv C_d/C_{AOD}I$ is the characteristic response time of the force-feedback system as we will see later. It is seen from eqn (8) that $\tilde{u}_{AOD}(\omega)$ and $\tilde{u}_d(\omega)$ are related by a phase difference of $\pi/2$,

$$\tilde{u}_d = -i\omega\tau_{PID}\tilde{u}_{AOD}. \quad (9)$$

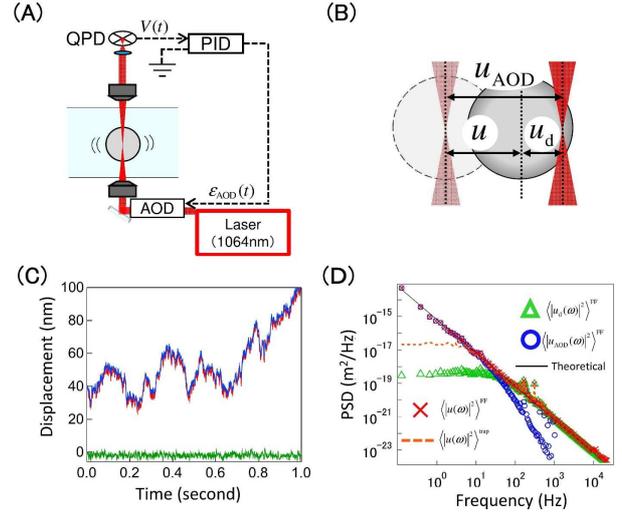

**Fig. 2** Force-feedback PMR. (A)(B) Schematic of the setup for force-feedback PMR. The QPD output $V(t)$ , which is proportional to displacement of the probe particle from the laser focus $u_d(t)$ , is fed to the input of the PID controller and its set point is grounded. The voltage signal $\varepsilon_{AOD}(t) = I\int V(t)dt$ generated in the PID controller is fed to the AOD, and controls the focus position of the laser, $u_{AOD}(t) = C_{AOD}\varepsilon_{AOD}(t)$ . (C) The displacement of the probe particle (melamine resin, $2a = 1$ μm), $u(t) = u_{AOD}(t) - u_d(t)$ (red line), $u_{AOD}(t)$ (blue line), and $u_d(t)$ (green line) measured by force-feedback PMR in 90% glycerol at $37°C$ . (D) PSD measured by force-feedback PMR $\langle|\tilde{u}(\omega)|^2\rangle^{FF}$ (red crosses) and by conventional MR $\langle|\tilde{u}(\omega)|^2\rangle^{trap}$ (orange broken line) in 90% glycerol. Green triangles and blue circles are $\langle|\tilde{u}_d(\omega)|^2\rangle^{FF}$ and $\langle|\tilde{u}_{AOD}(\omega)|^2\rangle^{FF}$ measured by force-feedback PMR respectively. While $\langle|\tilde{u}(\omega)|^2\rangle^{trap}$ in conventional MR is suppressed at low frequency, $\langle|\tilde{u}(\omega)|^2\rangle^{FF}$ agrees with the theoretical estimate $\langle|\tilde{u}(\omega)|^2\rangle = k_BT/3\pi a\omega^2\eta$ in 90% glycerol solution (black line). The measurement in 90% Glycerol using force-feedback PMR was carried out with $\tau_{PID} \sim 0.004$ s, $k = 5.1\times10^{-5}$ N/m.

The PSD of the probe displacement $u(t)$ is then described as

$$\langle|\tilde{u}(\omega)|^2\rangle^{FF} \equiv \langle|\mathcal{F}[u_{AOD}(t) - u_d(t)]|^2\rangle^{FF}$$
$$= \langle|\tilde{u}_d(\omega)|^2\rangle^{FF} + \langle|\tilde{u}_{AOD}(\omega)|^2\rangle^{FF} \quad (10)$$

where $\langle|\tilde{u}_d(\omega)|^2\rangle^{FF}$ , $\langle|\tilde{u}_{AOD}(\omega)|^2\rangle^{FF}$ and $\langle|\tilde{u}(\omega)|^2\rangle^{FF}$ are the

PSDs of $u_d(t)$, $u_{AOD}(t)$ and $u(t)$, and $\mathcal{F}[]$ indicates the Fourier transform, respectively. Here and hereafter, the superscript "FF" indicates the measured quantity under force-feedback control.

Performance of the force-feedback PMR was tested in 90% glycerol solution. Fig. 2C shows the measured displacements, $u_{AOD}(t)$, $u_d(t)$, and $u(t) = u_{AOD}(t) - u_d(t)$, respectively. In Fig. 2D, we show the PSD measured by conventional PMR, ($\langle|\tilde{u}(\omega)|^2\rangle^{trap}$: orange dash line), and those measured by force-feedback PMR, ($\langle|\tilde{u}(\omega)|^2\rangle^{FF}$: red crosses, $\langle|\tilde{u}_d(\omega)|^2\rangle^{FF}$: green triangles, $\langle|\tilde{u}_{AOD}(\omega)|^2\rangle^{FF}$: blue circles), respectively. The solid black line in the same graph indicates the theoretical estimate

$$\langle|\tilde{u}(\omega)|^2\rangle = k_B T / 3\pi a \omega^2 \eta, \quad (11)$$

where $\eta = 0.14$ Pa·s is the viscosity of the solution [34] and $a = 0.5$ μm is the radius of the probe particle (melamine particle, microParticles GmbH). In both conventional PMR and force-feedback PMR, the strength of the optical trap was set to $k = 5.1 \times 10^{-5}$ N/m.

In conventional PMR, the probe fluctuation $\langle|\tilde{u}(\omega)|^2\rangle^{trap}$ was suppressed by the optical trap below the frequency $f_c \equiv 1/(2\pi\tau_c) \sim 6$ Hz, as stated previously [23]. Here, $\tau_c = \gamma_0/k \sim 0.03$ s is the response time of the trapped particle, and $\gamma_0 = 6\pi\eta a$ is the friction coefficient of a probe in purely viscous material. On the other hand, in force-feedback PMR, the total probe fluctuation $\langle|\tilde{u}(\omega)|^2\rangle^{FF}$ completely agrees with the theoretical estimate, $\langle|\tilde{u}(\omega)|^2\rangle = k_B T / 3\pi a \omega^2 \eta$. This observation indicates that the force-feedback works as intended; probe fluctuations are not affected by the optical-trapping potential. Eqn (10) is confirmed by calculating PSDs of $u_d(t)$ and $u_{AOD}(t)$ separately. The total fluctuation $\langle|\tilde{u}(\omega)|^2\rangle^{FF}$ is mostly composed of $\langle|\tilde{u}_{AOD}(\omega)|^2\rangle^{FF}$ at frequencies below ~40 Hz. On the other hand, $\langle|\tilde{u}_d(\omega)|^2\rangle^{FF}$ takes over at frequencies higher than $1/2\pi\tau_{PID} \sim 40$ Hz since the AOD-controlled laser is unable to follow the probe fluctuation. This crossing-over behavior complies with the theoretical expectation which we obtain from eqns. (9) and (10),

$$\langle|\tilde{u}_d(\omega)|^2\rangle^{FF} = \omega^2 \tau_{PID}^2 \times \langle|\tilde{u}_{AOD}(\omega)|^2\rangle^{FF} = \frac{\omega^2 \tau_{PID}^2}{1 + \omega^2 \tau_{PID}^2} \langle|\tilde{u}(\omega)|^2\rangle^{FF}. \quad (12)$$

The crossover frequency ($\sim 40$ Hz) is consistent with the estimated response time of the feedback $\tau_{PID} \sim 0.004$ s (calculated from $C_{AOD} = 7.1 \times 10^{-6}$ m/V, $C_d = 1.4 \times 10^{-7}$ m/V, $I = 5$ s$^{-1}$).

In reality, there is an additional delay between $u_{AOD}(t)$ and $\varepsilon_{AOD}(t)$ since it takes $\tau_{AOD} \sim 10^{-5}$ s for the ultrasonic wave generated at the edge of the AOD to propagate and arrive at the center of the element where the laser light is deflected. $\varepsilon_{AOD}(t)$ is also delayed from the QPD output $V(t)$ by

$\tau_{contr} \sim 10^{-6}$ s. In total, $u_{AOD}(t)$ is delayed from $u_d(t)$ by $\tau \equiv \tau_{AOD} + \tau_{contr}$, as shown in Supplementary 1. Note that eqn (8)-(10) and (12) were obtained by neglecting this time delay, assuming $\tau \ll \tau_{PID}$. In order to achieve the expected performance of the force feedback, both $\tau_{PID}$ and $\tau$ have to be much smaller than $\tau_c \equiv \gamma_0/k$, i.e. $\tau \ll \tau_{PID} \ll \tau_c$. In the experiment shown in Fig. 2, these conditions were satisfied since $\tau_c$ is large in the highly viscous sample. When the sample is less viscous (e.g. water) or the trap stiffness is greater, $\tau_c$ will be decreased, by many orders of magnitude. Then, eqn (10) and (12) are not satisfied because $\tau \ll \tau_{PID} \ll \tau_c$ is not attained. Experimental tests and theoretical analysis of the feedback PMR under the condition $\omega \geq 1/\tau \sim 1/\tau_{PID}$ are given in Supplementary 2, and those for $\omega \leq 1/\tau_c < 1/\tau_{PID}$ are given in Supplementary 3.

### 1.2.2 Force-feedback AMR

For AMR, a sinusoidal force is applied to the probe particle by the drive laser ($\lambda = 1064$ nm) operated under force feedback. This is achieved in a manner similar to force-feedback PMR except that the sinusoidal signal $s(t) = L e^{-i\omega t}/C_d$ is fed to the set point of the PID controller instead it is grounded (Fig. 3A). For $\omega \ll 1/(\tau + \tau_{PID})$, the focus position of the drive laser is given as $u_d(t) = L e^{-i\omega t}$ by the feedback control. The optical-trapping force $F(t) = k_d u_d(t) = k_d L e^{-i\omega t}$ is independent of the stochastically fluctuating probe displacement $u(t)$ and, therefore, is well controlled. The displacement response of the probe particle to the applied force is detected with BFPI using a probe laser ($\lambda = 830$ nm) whose focus position is fixed (Fig. 3B).

The Langevin equation for the motion of the probe particle is then given by

$$k_p u(t) + \int_{-\infty}^t \gamma(t-t')\dot{u}(t')dt' = k_d u_d(t) + f(t), \quad (13)$$

where $k_p u(t)$ and $k_d u_d(t) = F(t)$ are the forces applied by the probe laser and the drive laser, respectively. The ensemble average of eqn (13) then yields the frequency-dependent response, which is written as

$$k_p \langle\hat{u}(\omega)\rangle^{FF} - i\omega\tilde{\gamma}\langle\hat{u}(\omega)\rangle^{FF} = k_d L. \quad (14)$$

The intrinsic response function $\alpha(\omega) = -1/[i\omega\tilde{\gamma}(\omega)]$ is given by

$$\alpha(\omega) = \frac{A_{FF}(\omega)}{1 - k_p A_{FF}(\omega)}, \quad (15)$$

where $A_{FF}(\omega)$ is the apparent response function $A_{FF}(\omega) \equiv \langle\hat{u}(\omega)\rangle^{FF}/k_d L$ under force feedback.

So far, we have neglected delay times, $\tau_{PID}$ and $\tau = \tau_{AOD} + \tau_{contr}$. This approximation holds if the force-feedback AMR is conducted limitedly at low frequencies where conventional AMR does not work well owing to the optical-trapping potential. Since $\tau + \tau_{\mathrm{PID}} \ll 1/\omega$ is satisfied in such a case, $\alpha(\omega)$ is obtained from eqn (15). When $\tau + \tau_{\mathrm{PID}} \geq 1/\omega$ , feedback does not achieve the expected response which we have assumed so far, *i.e.* $u_{\mathrm{d}}(t) = Le^{-i\omega t}$ and $F(t) = k_{\mathrm{d}}Le^{-i\omega t}$ . In this case, the actual $u_{\mathrm{d}}(t)$ [ $= u_{\mathrm{AOD}}(t) - u(t)$ ] must be estimated by considering how $u_{\mathrm{AOD}}(t)$ is delayed owing to $\tau_{\mathrm{PID}}$ and $\tau$ . Explicitly, it is made as

$$u_{\mathrm{AOD}}(t) = C_{\mathrm{AOD}} \, \varepsilon_{\mathrm{AOD}}(t - \tau_{\mathrm{PID}}) = (1/\tau_{\mathrm{PID}}) \int_{-\infty}^{t-\tau} \{u_{\mathrm{d}}(t') - Le^{-i\omega t'}\} \, dt' \ . \quad (16)$$

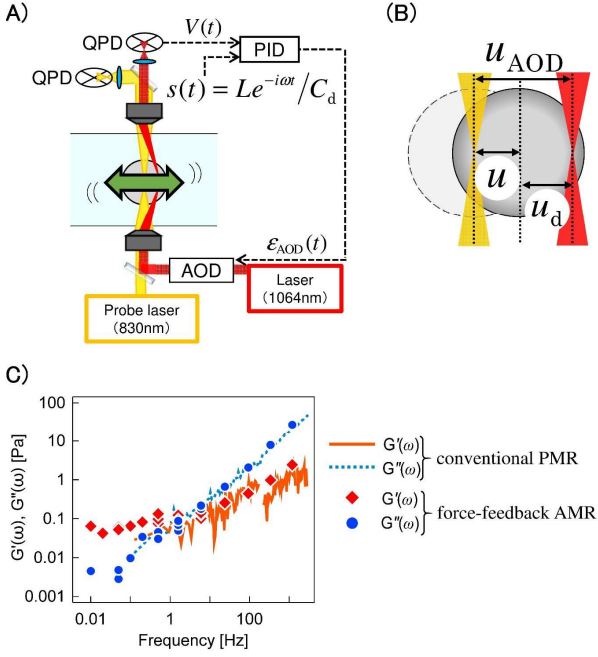

**Fig. 3** Force-feedback AMR (A) (B) Schematic of the setup for force-feedback AMR. An AOD-controlled drive laser ( $\lambda = 1064$ nm ) applied a sinusoidal force to the probe particle, and a probe laser ( $\lambda = 830$ nm ) detected the displacement of the probe $u$ . By feeding the sinusoidal signal $s(t) = Le^{-i\omega t}/C_{\mathrm{d}}$ to the set point of the PID controller, the laser focus of drive laser oscillates around the center of fluctuating probe particle as $u_{\mathrm{d}}(t) = u_{\mathrm{AOD}}(t) - u(t) = Le^{-i\omega t}$ (C) Shear viscoelastic modulus $G(\omega)$ of a 0.6 mg/mL entangled F-actin gel. Closed symbols were measured with force-feedback AMR and curves were measured with conventional PMR. Red and blue symbols are the real [ $G'(\omega)$ ] and imaginary [ $G''(\omega)$ ] parts of $G(\omega)$ , respectively. For force-feedback AMR, $\tau_{\mathrm{PID}} \sim 2.4 \times 10^{-5}$ s , $k_{\mathrm{d}} \sim 1.1 \times 10^{-5}$ N/m and $k_{\mathrm{p}} \sim 1.6 \times 10^{-6}$ N/m .

By incorporating $u_{\mathrm{AOD}}(t) = \hat{u}_{\mathrm{AOD}}(\omega)e^{-i\omega t}$ , $u_{\mathrm{d}}(t) = \hat{u}_{\mathrm{d}}(\omega)e^{-i\omega t}$ into eqn (16), and using $u(t) = u_{\mathrm{AOD}}(t) - u_{\mathrm{d}}(t)$ , we obtain

$$\langle \hat{u}_{\mathrm{d}}(\omega) \rangle = \frac{-i\omega\tau_{\mathrm{PID}} \langle \hat{u}(\omega) \rangle^{\mathrm{FF}} + Le^{i\omega\tau}}{i\omega\tau_{\mathrm{PID}} + e^{i\omega\tau}} \ . \quad (17)$$

By incorporating eqn (17) into (13), and noting $\hat{F}(\omega) = k_{\mathrm{d}}\hat{u}_{\mathrm{d}}\omega \neq k_{\mathrm{d}}L$ , the intrinsic response $\alpha(\omega) = -1/[i\omega\hat{\gamma}(\omega)]$ is obtained as,

$$\alpha(\omega) = \frac{A_{\mathrm{FF}}(\omega)(e^{i\omega\tau} + i\omega\tau_{\mathrm{PID}})}{e^{i\omega\tau}(1 - A_{\mathrm{FF}}(\omega)k_{\mathrm{p}}) - i\omega\tau_{\mathrm{PID}} A_{\mathrm{FF}}(\omega)(k_{\mathrm{d}} + k_{\mathrm{p}})} \ . \quad (18)$$

The force-feedback AMR was conducted in a 0.6 mg/mL entangled F-actin gel. $\alpha(\omega)$ obtained using eqn (18) was substituted into eqn (3) to obtain complex shear modulus $G(\omega)$ . Results are shown in Fig. 3C (real part $G'(\omega)$ : solid red diamonds and imaginary part $G''(\omega)$ : solid blue circles). Conventional PMR shows similar results [ $G'(\omega)$ : orange solid curves and $G''(\omega)$ : light blue broken curves], but is limited in its frequency range $f \geq 1$ Hz . The precision of conventional PMR is decreased since errors enter due to the optical-trapping potential and low-frequency noise. On the other hand, force-feedback AMR provides $G(\omega)$ at lower frequencies where it is challenging to measure with conventional methods [23, 24, 28, 35]

## 2. Nonlinear Dual Feedback MR

### 2.1 Dual feedback system

Force-feedback MR cannot be precisely conducted when the AOD-controlled laser ( $\lambda = 1064$ nm ) moves away from the optical axis of the objective lens while it is accompanying the fluctuating probe particle. As shown in Supplementary 4, an offset in QPD output and an error in the calibration factor $C_{\mathrm{d}}$ appear when the laser is far ($\sim$10 μm) from the optical axis. The force-feedback MR experiments presented in prior sections were therefore conducted in a highly viscous sample in which the thermal Brownian motion was sufficiently reduced. Additionally, the force feedback follows the probe particle only in lateral directions, but probe fluctuations along the optical axis can also introduce significant errors due to the sensitivity of BFPI [36]. In this study, in order to track a largely fluctuating probe [37], another feedback control mechanism referred to as *stage feedback* [33] was introduced in addition to the force-feedback.

As detailed in our prior study [33], stage feedback was applied in three dimensions (3D) by controlling the piezo stage on which a sample chamber is placed (Fig. 4A) [33]. The probe displacements in lateral ($x$-, $y$-) directions were measured by BFPI using a fixed probe laser ($\lambda = 830$ nm); displacements in the axial ($z$-) direction were measured by analyzing the pattern of the microscope image of the probe particle. The piezo stage was then controlled using PID feedback. Since the response of the piezo stage is slow due to its inertia, the feedback-response time was set much larger than the force feedback. This stage feedback can be performed within the travel range of the piezo stage, ~200 μm in this study. The piezo stage then tracks the slow/large fluctuations of the probe $u_s(t)$ while the fast/small fluctuations are detected by the BFPI using the probe laser $u_p(t)$. The total displacement of the probe $u(t)$ is found using $u(t) = u_p(t) + u_s(t)$, where $u_s(t)$ is the displacement of the piezo stage and $u_p(t)$ is the distance between the probe particle and the focus of the probe laser (Fig. 4B).

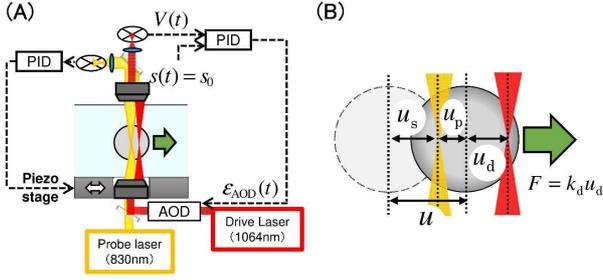

**Fig. 4** Force-clamp MR under dual feedback. (A) Schematic of the setup for force-clamp MR. Both force-feedback control (right loop) and stage feedback control (left loop) are carried out simultaneously. For the force feedback, a constant offset signal $s(t) = s_0$ is fed to the set point of the PID controller to keep a constant distance between the center of the fluctuating probe particle and the drive laser, $u_d$. For stage feedback, the displacement of the piezo stage $u_s$ is controlled to locate the probe particle around the laser focus of the fixed probe laser. If sinusoidal signal $s(t) = Le^{-i\alpha t}/C_d$ is fed to the set point of the PID controller for the force feedback, dual feedback AMR is carried out. (B) $u_d$ is maintained by force-feedback control. A stable, constant force $F = k_d u_d$ is applied to the probe in one direction. Total displacement of the probe $u$ is obtained by summing $u_p$ and $u_s$.

In dual feedback linear AMR, a well-controlled force will be applied to the probe particle by the drive laser $F(t) \sim k_d Le^{-i\alpha t}$; the response can then be measured in either $u_s(t)$ or $u_p(t)$ depending on the applied frequency because they are correlated via the feedback. Note that the probe laser is operated at low power since it is used solely for detecting the position of the particle. But even if the force applied by the probe laser to the probe particle is not negligible, it can be corrected following existing procedures [33]. When the feedback-response times for the two target apparatuses, the piezo stage and the AOD, are similar, their feedback controls would destabilize since they interfere with each other. This did not happen in our setup since, in ordinary situations, the feedback-response time of the force feedback is much faster ( $> 10^3$ times) than that of stage feedback.

## 2.2 Force-clamp MR under dual feedback

For nonlinear MR, a colloidal particle embedded in a soft material must be forced beyond the linear-response regime. The applied force will then induce vigorous probe fluctuations which must be measured with high spatiotemporal resolution. One of the technical challenges here is again to apply a well-controlled force on a vigorously fluctuating probe particle. This is achieved by using a constant setpoint $s(t) = s_0 \neq 0$ for the force feedback applied by the drive laser. This method, referred to as the force-clamp mode, applies a constant force to the probe particle. Although the probe particle will drift over large distances, BFPI with the probe laser can be achieved precisely by introducing stage feedback, keeping the particle close to the optical axis. (Fig. 4A).

When the setpoint for the force feedback is constant, $s(t) = s_0$, the control signal $\varepsilon_{AOD}(t)$ produced by the PID controller becomes $\varepsilon_{AOD}(t) = I \int V(t) - s_0 \, dt$. The distance between the center of the probe and focus of the drive laser is kept constant at $u_d = C_d s_0$ (Fig. 4B), and the applied force is described as $F = k_d u_d$. The displacement of the probe $u(t)$ is obtained from the sum of the displacement of the piezo stage $u_s(t)$ and the distance between the probe particle and the focus of the probe laser $u_p(t)$, $u(t) = u_p(t) + u_s(t)$. In addition to the information found in linear AMR and PMR experiments, note that this method explicitly allows for the extraction of an averaged viscosity of a sample $\eta$ in the long time limit, as given by Stokes relation $\eta = F/6\pi a v$, where $a$ is the radius of probe particles and $v$ is the velocity of probe particles.

## 2.3 Application to sparsely crosslinked F-actin gels

We now demonstrate the potential of force-clamp MR as a novel tool to investigate the nonlinear mechanical response of soft matter. A $2a = 2$ μm colloidal particle (Silica, Polysciences) was pulled in force-clamp mode through F-actin gels that were sparsely crosslinked with heavy meromyosin (F-actin 1.3 mg/mL and HMM 0.04 mg/mL). Without crosslinking, even a minimal force (< pN) can move the probes smoothly as if they were dispersed in viscous fluids. The restructuring of the entangled network, which will occur due to reptation [38], causes this less striking response. Crosslinks were thus introduced to suppress the spontaneous restructure of the gel.

Fig 5A shows the displacements of the probe particles in the direction of the force. The probe particles were trapped in the surrounding gel [16] when the applied force is small (*e.g.* $F = 1.0$ pN, yellow line and $F = 2.5$ pN, green line), guaranteeing that thermal reptation does not occur. Likewise, $\text{PSD} \cdot \omega / 2k_B T$ was not affected by the application of such small forces (Fig. 5B). At $F = 3.4$ pN, however, some probe particles started to move with intermittent jumps (blue lines in Fig. 5A which correspond to blue broken curve in Fig. 5B). When these intermittent jumps were not observed, the PSD slightly decreased at low frequencies (Fig. 5B, Blue solid curve), which is consistent with the stress stiffening of cytoskeletal gels [4, 11]. All probe particles experienced intermittent jumps when the applied force was increased further ($F$ = 4.3 pN red curve). The directed movements via intermittent jumps increased fluctuations at low frequencies whereas fluctuations at high frequencies were not changed (Fig. 5B). Similar behaviour has been frequently observed in various non-equilibrium systems [28, 39]; the increased fluctuations at low frequencies can be attributed to non-thermal fluctuations generated by energy input, provided here by the drive laser (broken curves in Fig. 5B).

Dynamics of the stochastic jumps were investigated with the probability distribution $P_{wtd}(t_w)$ of the waiting times $t_w$ between consecutive jumps. Jump events in the trajectory of the probe (Fig. 5C) were detected using the step detection algorithm. Step-detection algorithms have been established in single molecule studies and usually provide a better guess for when step events occur if the step size $\Delta x$ is *a priori* known to be a constant. Whereas many molecular motors perform steps with a constant stride, the step size of the hopping here distributes owing to the complexity of the network. Therefore, we chose the algorithm in which the step size for each jump is an adjustable parameter [40]. $P_{wtd}(t_w)$ of the forced jumps is shown in Fig 5D. The results were fitted well with an exponential function, $P_{wtd}(t_w) = k\exp(-kt_w)$, suggesting that each jump occurs following Poissonian statistics. Therefore, our experimental results indicate that the forced probe particles displayed Markov jumps that occurred randomly in time.

In addition, the dynamics of the forced probe particle were investigated with another approach that does not need to detect when the actual steps occur. We calculated the probability distribution $P(\Delta u, \Delta t)$ of the probe displacements $\Delta u$ in the direction of the applied force that occurred during a lag time $\Delta t$, referred to as van Hove distributions. For weak forcing ($F \leq 2.5$ pN), the shape of $P(\Delta u, \Delta t)$ did not evolve with $\Delta t$ and remained Gaussian (Fig. 6A). For stronger forcing ($F$ = 4.3 pN, Fig. 6B), the distribution function was close to Gaussian only when the lag time $\Delta t$ was small. As the lag time $\Delta t$ increases, the tail of the distribution extends in the direction of the force ($\Delta u > 0$) whereas the distribution in the opposite direction is hardly affected, remaining Gaussian.

Non-Gaussian tails have been frequently observed when a probe exhibits rare but large jumps [37]. In such a case, the area $S(\Delta t)$ exceeding the central Gaussian distribution (yellow region in Fig. 6B) indicates the probability that at least one jump occurred in $\Delta t$. The distribution of thermal fluctuations was estimated by fitting the Gaussian function to the central portion of the van Hove distribution, as shown in Fig. 6B. By subtracting the integrated probability of thermal fluctuations from the total, the area of the yellow region [ $S(\Delta t)$ ] in Fig. 6B was obtained. As shown in Fig. 6C, it is seen that $S(\Delta t)$ evolves linearly with $\Delta t$. Note that $S(\Delta t)$ is related to $P_{wtd}(t_w)$ by $S(\Delta t) = k\int_0^\infty dt_0 \int_{t_0}^{t_0 + \Delta t} P_{wtd}(t_w)dt_w$, under the assumption that jump events are independent as schematically shown in Fig. 6D. $S(\Delta t)$ can then be expressed as $S(\Delta t) = -\exp(-k\Delta t)+1$, consistent with the experimental result shown in Fig. 6C when $\Delta t$ is small. The consistency of two independent analyses confirmed that a probe particle subjected to optical-trapping force in a cytoskeletal network undergoes Markovian jumps.

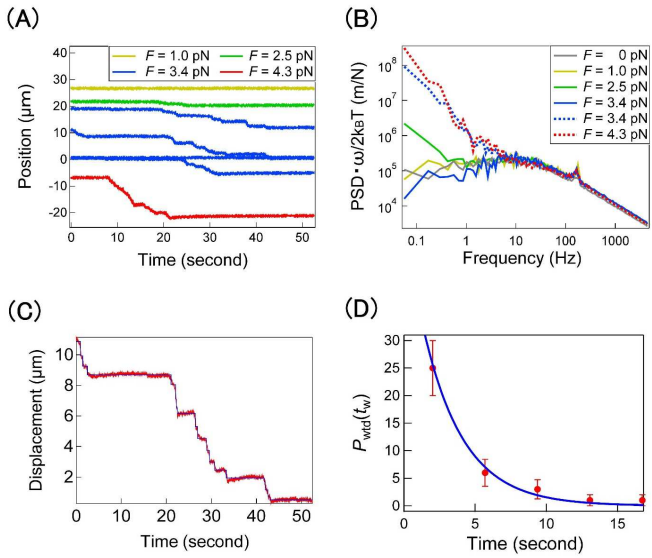

**Fig. 5** (A) Probe movements in a crosslinked F-actin gel under constant forcing ($F$), shown as a function of time. Nonlinear force-clamp MR was conducted with $\tau_{PID} \sim 7.7 \times 10^{-6}$ s, $k_d \sim 1.7 \times 10^{-5}$ N/m and $k_p \sim 1.5 \times 10^{-5}$ N/m. (B) $PSD \cdot \omega / 2k_B T$ obtained by force-clamp MR. When the applied force $F$ was small, the probe fluctuations are similar to those measured without forcing (grey solid line). Intermittent jumps were observed when the applied force was increased ($F$ = 3.4 pN: blue broken line, $F$ = 4.3 pN: pink broken line). These also appeared in the PSD as enhanced non-thermal fluctuations at low frequencies. (C) Red line shows probe movements in crosslinked F-actin gel under constant forcing 4.3 pN, observed with a nonlinear force-clamp. Blue line is the fit used to extract jumps of the probe using a step detection algorithm. (D) Red circles show the probability distribution $P_{wtd}(t_w)$ of the waiting times between consecutive jumps. The results were fit by an exponential function, $P_{wtd}(t_w) = k\exp(-kt_w)$ (blue curve).

## 3. Discussion

We developed feedback MR as a means to investigate the linear and nonlinear response of soft materials at the mesoscale (nm~μm). The advantages of the techniques compared with conventional MR were raised in prior sections. Here, we discuss the implications of the experiments shown in the last section (nonlinear MR in crosslinked actin gels) to highlight the potential of the technique.

In order to have insight into the observed probe dynamics, it is necessary to estimate the characteristic lengths of our F-actin sample. The mesh size $\xi$ of the semi-flexible network is obtained from the length density $\rho$ (= 24.5 μm$^{-2}$) of the filaments as $\xi \sim 1/\sqrt{\rho} \sim 140$ nm [38]. The average distance $l_c$ between crosslinks along each F-actin filament is estimated as

$$l_c \sim \left(6\rho k_B T l_p^2 / G\right)^{1/3}, \quad (19)$$

where $l_p = 10$ μm is the persistent length of actin filaments without phalloidin labelling. $G$ is the elastic plateau modulus of F-actin / HMM gels that were not subjected to external forces [41]. We then obtain $l_c \sim 7$ μm whereas the average distance between nearest crosslinks is $l_n \sim 0.5$ μm (Fig. 6E). These characteristic length scales of the F-actin gel (mesh size $\xi$, persistent length $l_p$, the crosslink distance $l_c$, etc) are similar to the size of the probe particle ($a = 1$ μm) in order of magnitude, and therefore could profoundly affect the probe dynamics.

Without external forcing, probe particles are deeply constrained in the potential wells formed by the elastic microenvironments of the crosslinked gel. Therefore, thermal fluctuations could reflect merely the bottom curvature of the potential. The Gaussian nature (Fig. 6A) of the distribution under $F = 2.5$ pN implies that the medium surrounding the probe is regarded as a homogeneous continuum, as far as linear MR is concerned. This is likely because our probes are constrained in the network with $\xi$ smaller than the probe size. On the other hand, the jump process observed under nonlinear forcing ( $F \geq 3.4$ pN ) may reflect the whole depth of the potential, associated with sparse crosslinks rather than the mesh of the network. Even if $l_n < a$, position of filaments and crosslinks can rearrange to allow probe hopping. It is not necessary to break the actin filaments or the HMM crosslinks.

It has been reported that probe particles in dilute ($\xi \sim a$) and non-crosslinked F-actin networks jump intermittently and spontaneously, in the absence of any external force [16, 42]. Such thermal jumps also show non-Gaussian dynamics with side tails in the van Hove distributions. However, the waiting-time distribution of the thermal hopping followed a power-law function, $P_{wtd}(t) \propto 1/t^\alpha$ ( $1 < \alpha < 2$ ) [16, 42], in contrast to the forced jumps of the probe observed in this study. It was reported that the observed power-law distribution is consistent with the theoretical model for anomalous diffusion: a continuous time random walk (CTRW) whose waiting times have a distribution with a power-law decaying tail [43]. Bouchaud's trap model [44-46] links the power-law distribution of waiting times to the heterogeneity of microenvironments. In the theory, glassy dynamics are attributed to the probability density of microenvironments $\rho(E)$ having a potential depth $E$ [44]. Since thermal probes are trapped longer in deeper potentials following Boltzmann's statistics, a power-law distribution of waiting times $P_{wtd}(t)$ is observed. These prior reports, both experiments and theories, thus indicate the presence of mechanically heterogeneous microenvironments that frequently show up in soft glassy materials [16, 20, 42].

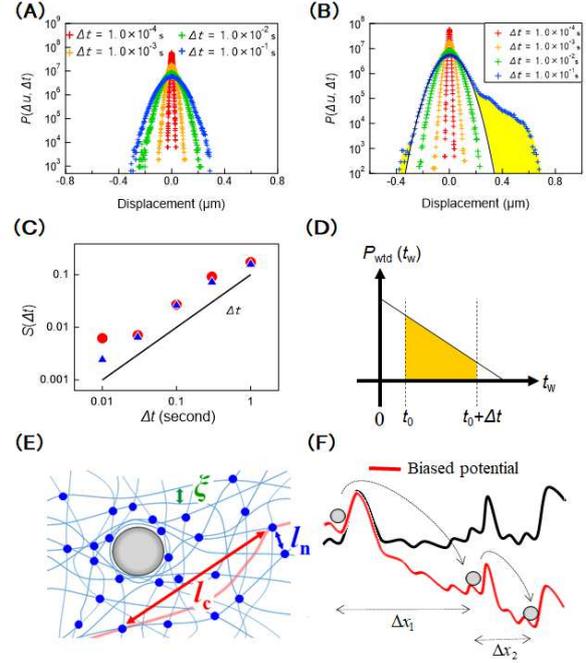

**Fig. 6** (A) The probability distribution $P(\Delta u, \Delta t)$ of probe displacements $\Delta u$ in duration $\Delta t$ , measured with force-clamp MR under dual feedback. A constant force ($F = 2.5$ pN) was applied to the probe in positive direction. (B) $P(\Delta u, \Delta t)$ measured at $F = 4.3$pN. The central portion of the distribution ( $\Delta t = 0.1$ s ) was fit with a Gaussian (black curve). The area of the yellow region $S(\Delta t)$ was obtained by subtracting the Gaussian from the total fluctuations. (C) $S(\Delta t)$ plotted as a function of the lag time $0.01$ s $\leq \Delta t \leq 1$ s . (D) A schematic describing the relation between $S(\Delta t)$ and $P_{wtd}(t)$ . Provided that the last step took place at $t = 0$, the area colored in orange indicates $S(\Delta t)$. (E) Schematic describing the characteristic lengths of crosslinked actin: mesh size $\xi$ (~ 140 nm), distance between crosslinks along the same filament $l_c$ (~ 7 μm), the distance between nearest crosslinks $l_n$ (~ 0.5 μm). Blue circles are crosslinks. (F) Schematic describing hopping of probe particle to neighbouring potential wells which are biased by the applied force (red curve). Hopping does not show glassy heterogeneous dynamics since the tracer bead is not trapped in shallow sub-basins in the original potential wells (black curve).

The exponential decay of $P_{wtd}(t)$ supports that the forced probe hopping between microenvironments follows a Markov process, and that the potential depths provided by different microenvironment are narrowly distributed. In non-crosslinked gels used in the prior study, the energy landscape contains small basins whose depth is largely distributed. The probe particle was temporarily trapped in such basins. In the sparsely cross-linked actin gels prepared in this study, we expect that crosslinks would create global wells which are much deeper than $k_B T$, in addition to the small subbasins as shown in Fig. 6F. Because $l_n < a$, we believe that many crosslinks are involved in the global well (Fig. 6E). A probe particle needs to squeeze out from many crosslinks encircling it to hop to a neighbouring microenvironment. From the statistical reason, it is then reasonable to expect that the threshold energy (depth of the global potential) is large, but may not be largely distributed (Fig 6F). Note that the external force effectively decreases the potential $E$ by a margin much larger than the thermal energy ( $F\Delta x \gg k_B T$ ). Under such strong forcing, small or intermediate sub-basins in the energy landscape will not trap the tracer beads (Fig. 6F).

As seen in Supplementary 5, the step size $\Delta x$ of probe hopping was largely distributed as expected for a specimen with heterogeneous microenvironments. However, the work done by the external force during the step ( $F\Delta x$ ) and the waiting time before the step ( $t_w$ ) did not show correlation. Therefore, the large distribution of $\Delta x$ and homogeneous hopping dynamics coexist in our specimen. The hopping dynamics became Poissonian as if they are occurring in homogeneous medium although probe particles are actually dispersed in heterogeneous microenvironments.

## 4. Materials

G-actin was prepared from rabbit skeletal muscle according to standard protocols cite [47] and was stored at -80 °C in G-buffer [2 mM tris-Cl, 0.2 mM $CaCl_2$, 0.5 mM DTT, and 0.2 mM ATP (pH 7.5)]. G-actin was diluted into F-buffer [1 mM $Na_2ATP$, 2 mM Hepes, 1 mM EGTA, 2 mM $MgCl_2$, and 50 mM KCl (pH 7.5)] to initiate actin polymerization. To prepare an entangled F-actin gel, F-actin solution (0.6 mg/ml) including polystyrene beads (Polysciences Inc., $2a = 1\,\mu m$ ) was directly infused into sample chambers. To prepare a crosslinked F-actin gel, G-actin (1.3 mg/ml) and HMM (Cytoskeleton Inc., 0.04 mg/ml) and polystyrene beads were mixed together and then infused into sample chambers. In both entangled and crosslinked actin gels, polymerization occurred at room temperature for about 1 hour. The thermal equilibrium sample was then measured by feedback PMR with the following calibration values: $k_t = k_p = 4.8 \times 10^{-6}$ N/m and $\tau = 0.024$ s.

Aqueous solution of 90% glycerol was prepared with the probe particles (diameter, $2a = 1\,\mu m$ ; refractive index, 1.68; polydispersity, <5%; microParticles GmbH). This thermal equilibrium sample was measured by force-feedback PMR with the following calibration values: $k = 5.1 \times 10^{-5}$ N/m and $\tau_{PID} \sim 0.004$ s (calculated from $C_{AOD} = 7.1 \times 10^{-6}$ m/V, $C_d = 1.4 \times 10^{-7}$ m/V, $I = 5$ s$^{-1}$ ).

## 5. Conclusion

In this study, we developed a technique to conduct optical-trap and laser-interferometry based MR under force-feedback control, where the trapping laser rapidly follows the probe fluctuations. Since the feedback enables to apply a well-controlled optical-trapping force to a probe particle, it is suitable for investigating slow response in soft materials. However, the force feedback loses its accuracy in applying a well-controlled force when the probe particle deviates strongly from the optical axis for the optical-trapping system. Therefore, when a probe particle greatly moves during each experiment, another feedback control referred to as the stage-feedback was performed in conjunction with the force feedback technique [33]. With this dual feedback technique, a vigorously fluctuating particle can be tracked close to the optical axis, which permits precise control of the optical-trapping force. Depending on the sample of interest and the purpose of the MR experiment, these techniques can be used separately or in combination in the dual feedback mode.

After validating the advantages of the developed technique, we conducted nonlinear MR under dual-feedback control. By applying a constant force to a probe particle embedded in a crosslinked F-actin gel beyond its linear response regime, we observed hopping of the probe to neighbouring microenvironments. Although thermal hopping in similar cytoskeletal networks (actin gels) reportedly showed power-law heterogeneous dynamics, the hopping of a forced probe was found to be Markovian [16].

Living organisms are made of soft matter (*e.g.* actin and myosin used in this study) and they are commonly driven far from equilibrium. In living cells, forces are spontaneously generated by molecular motors. Organelles and vesicles are transported by motors along cytoskeletal filaments. Owing to the nonlinear response characteristics of soft matter in the cytoplasm, such mechanical perturbations profoundly modulate the mechanics of living systems, as observed *in vitro* [28, 48] and *in vivo* [33, 49]. Understanding the linear and nonlinear response of biological soft matter to physiological forcing is thus the key to elucidating the intriguing mechanics of living systems [49, 50]. However, motor-generated forces are largely stochastic and therefore hard to control artificially. The feedback techniques presented in this study enable the application of well-controlled localized forces to probes embedded in a soft medium, making them appropriate for the mechanistic investigation of biological soft matter systems. The experiment presented here has demonstrated its potential, by revealing that homogeneous dynamics emerge under nonlinear forcing. As such, the results of linear and nonlinear MR carried out on biological soft matter may have abundant implications for our understanding of the mechanics of biological systems.

## Conflicts of interest

There are no conflicts to declare.

## Acknowledgements

This work was supported by JSPS KAKENHI Grant Number JP21H01048, JP20H05536, JP20H00128.

## Notes and references

1. S. Glasstone, K. J. Laidler and H. Eyring, The theory of rate processes; the kinetics of chemical reactions, viscosity, diffusion and electrochemical phenomena, McGraw-Hill Book Company, inc., New York; London,, 1st edn., 1941.
2. C. Kittel, Introduction to solid state physics, Wiley, New York, 6th edn., 1986.
3. M. Born and K. Huang, Dynamical theory of crystal lattices, Clarendon Press; Oxford University Press, Oxford, New York, 1985.
4. D. A. Head, E. Ikebe, A. Nakamasu, P. Zhang, L. G. Villaruz, S. Kinoshita, S. Ando and D. Mizuno, Phys Rev E Stat Nonlin Soft Matter Phys, 2014, 89, 042711.
5. F. Gittes, B. Schnurr, P. D. Olmsted, F. C. MacKintosh and C. F. Schmidt, Phys Rev Lett, 1997, 79, 3286-3289.
6. T. G. Mason and D. A. Weitz, Phys Rev Lett, 1995, 75, 2770-2773.
7. S. Jabbari-Farouji, M. Atakhorrami, D. Mizuno, E. Eiser, G. H. Wegdam, F. C. Mackintosh, D. Bonn and C. F. Schmidt, Phys Rev E Stat Nonlin Soft Matter Phys, 2008, 78, 061402.
8. S. Rafai, L. Jibuti and P. Peyla, Phys Rev Lett, 2010, 104, 098102.
9. M. Guvendiren, H. D. Lu and J. A. Burdick, Soft Matter, 2012, 8, 260-272.
10. P. A. Janmey, U. Euteneuer, P. Traub and M. Schliwa, J Cell Biol, 1991, 113, 155-160.
11. M. L. Gardel, J. H. Shin, F. C. MacKintosh, L. Mahadevan, P. Matsudaira and D. A. Weitz, Science, 2004, 304, 1301-1305.
12. N. J. Wagner and J. F. Brady, Physics Today, 2009, 62, 27-32.
13. P. Sehgal, M. Ramaswamy, I. Cohen and B. J. Kirby, Phys Rev Lett, 2019, 123, 128001.
14. E. R. Weeks, J. C. Crocker, A. C. Levitt, A. Schofield and D. A. Weitz, Science, 2000, 287, 627-631.
15. M. Atakhorrami, G. H. Koenderink, J. F. Palierne, F. C. Mackintosh and C. F. Schmidt, Phys Rev Lett, 2014, 112, 088101.
16. I. Y. Wong, M. L. Gardel, D. R. Reichman, E. R. Weeks, M. T. Valentine, A. R. Bausch and D. A. Weitz, Phys Rev Lett, 2004, 92, 178101.
17. B. Wang, J. Kuo, S. C. Bae and S. Granick, Nat Mater, 2012, 11, 481-485.
18. N. Nijenhuis, D. Mizuno, C. F. Schmidt, H. Vink and J. A. Spaan, Biomacromolecules, 2008, 9, 2390-2398.
19. N. Nijenhuis, D. Mizuno, J. A. Spaan and C. F. Schmidt, J R Soc Interface, 2012, 9, 1733-1744.
20. P. Sollich, F. Lequeux, P. Hebraud and M. E. Cates, Phys Rev Lett, 1997, 78, 2020-2023.
21. F. Gittes and C. F. Schmidt, Methods Cell Biol, 1998, 55, 129-156.
22. T. G. Mason, K. Ganesan, J. H. vanZanten, D. Wirtz and S. C. Kuo, Phys Rev Lett, 1997, 79, 3282-3285.
23. D. Mizuno, D. A. Head, F. C. MacKintosh and C. F. Schmidt, Macromolecules, 2008, 41, 7194-7202.
24. L. A. Hough and H. D. Ou-Yang, Phys Rev E Stat Nonlin Soft Matter Phys, 2002, 65, 021906.
25. D. Mizuno, Y. Kimura and R. Hayakawa, Phys Rev Lett, 2001, 87, 088104.
26. B. Schnurr, F. Gittes, F. C. MacKintosh and C. F. Schmidt, Macromolecules, 1997, 30, 7781-7792.
27. D. Mizuno, Y. Kimura and R. Hayakawa, Langmuir, 2000, 16, 9547-9554.
28. D. Mizuno, C. Tardin, C. F. Schmidt and F. C. Mackintosh, Science, 2007, 315, 370-373.
29. S. Jabbari-Farouji, D. Mizuno, D. Derks, G. H. Wegdam, F. C. MacKintosh, C. F. Schmidt and D. Bonn, Epl-Europhys Lett, 2008, 84.
30. L. G. Wilson, A. W. Harrison, W. C. K. Poon and A. M. Puertas, Epl-Europhys Lett, 2011, 93.
31. N. Senbil, M. Gruber, C. Zhang, M. Fuchs and F. Scheffold, Phys Rev Lett, 2019, 122, 108002.
32. F. Gittes and C. F. Schmidt, Opt Lett, 1998, 23, 7-9.
33. K. Nishizawa, M. Bremerich, H. Ayade, C. F. Schmidt, T. Ariga and D. Mizuno, Sci Adv, 2017, 3, e1700318.
34. A. Glycerine Producers, Physical properties of glycerine and its solutions, Glycerine Producers' Association, 1963.
35. H. Turlier, D. A. Fedosov, B. Audoly, T. Auth, N. S. Gov, C. Sykes, J. F. Joanny, G. Gompper and T. Betz, Nature Physics, 2016, 12, 513-519.
36. Y. Sugino, M. Ikenaga and D. Mizuno, Applied Sciences-Basel, 2020, 10, 4970.
37. T. Kurihara, M. Aridome, H. Ayade, I. Zaid and D. Mizuno, Phys Rev E, 2017, 95, 030601.
38. M. Doi and S. F. Edwards, The theory of polymer dynamics, Clarendon Press; Oxford University Press, Oxford, New York, 1986.
39. T. Ariga, M. Tomishige and D. Mizuno, Phys Rev Lett, 2018, 121, 218101.
40. J. W. J. Kerssemakers, E. L. Munteanu, L. Laan, T. L. Noetzel, M. E. Janson and M. Dogterom, Nature, 2006, 442, 709-712.
41. F. C. MacKintosh, J. Kas and P. A. Janmey, Phys Rev Lett, 1995, 75, 4425-4428.
42. B. Wang, S. M. Anthony, S. C. Bae and S. Granick, Proc Natl Acad Sci U S A, 2009, 106, 15160-15164.
43. R. Metzler and J. Klafter, Phys Rep, 2000, 339, 1-77.
44. C. Monthus and J. P. Bouchaud, J Phys a-Math Gen, 1996, 29, 3847-3869.
45. E. M. Bertin and J. P. Bouchaud, Phys Rev E Stat Nonlin Soft Matter Phys, 2003, 67, 026128.
46. J. P. Bouchaud, J Phys I, 1992, 2, 1705-1713.
47. D. W. Frederiksen and L. W. Cunningham, methods enzymol, 1982, 85.
48. C. P. Brangwynne, G. H. Koenderink, F. C. Mackintosh and D. A. Weitz, Phys Rev Lett, 2008, 100, 118104.
49. K. Nishizawa, K. Fujiwara, M. Ikenaga, N. Nakajo, M. Yanagisawa and D. Mizuno, Sci Rep-Uk, 2017, 7, 15143.
50. D. Humphrey, C. Duggan, D. Saha, D. Smith and J. Kas, Nature, 2002, 416, 413-416.

# Supplementary

## 1: Evaluation of the delay times: $\tau_{\mathrm{AOD}}$ and $\tau_{\mathrm{contr}}$

To accurately measure the displacement response of the probe to the optical-trapping force, we checked the time delays of all instruments. In the range of frequencies where MR was performed (up to 100 kHz in this study), time delays produced by many of the instruments were negligible, except for $\tau_{\mathrm{AOD}}$ and $\tau_{\mathrm{contr}}$ which were generated by the AOD (Acousto-Optic deflector) and the PID-feedback controller, respectively. In the MR experiments presented in the main text, the force feedback was not operated too fast; $\tau_{\mathrm{AOD}}$ and $\tau_{\mathrm{contr}}$ were sufficiently smaller than the characteristic response time of the force feedback $\tau_{\mathrm{PID}} \equiv C_{\mathrm{d}} / C_{\mathrm{AOD}} I$ (see the main text for definitions of $C_{\mathrm{d}}$, $C_{\mathrm{AOD}}$ and $I$). In the following sections, we will investigate the response of our experimental set up under the faster force feedback where $\tau_{\mathrm{contr}}$ and $\tau_{\mathrm{AOD}}$ are not negligible compared to $\tau_{\mathrm{PID}}$. In order to proceed with the investigation, the delay times, $\tau_{\mathrm{AOD}}$ and $\tau_{\mathrm{contr}}$ were measured as written below.

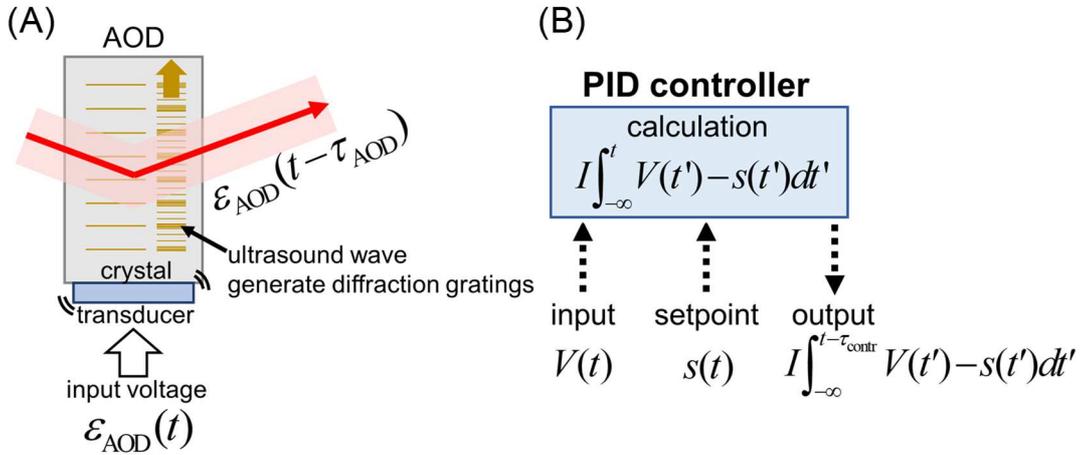

**Fig. S1** Response-time delay of AOD and PID controller. (A) Inside of the AOD, transducer generates ultrasound wave to form diffraction grating, and laser is diffracted by this grating. Thereby, the laser diffraction delays from the AOD control signal $\varepsilon_{\mathrm{AOD}}(t)$ by the propagation time $\tau_{\mathrm{AOD}}$ of the sound wave in the AOD. (B) The QPD signal $V(t)$ is input to PID controller. The controller generates output signal which will bring QPD signal $V(t)$ closer to "setpoint" $s(t)$ via feedback (*e.g.* $s(t) = 0$ for force-feedback PMR). Here, the output signal is given by the integral of the difference between $V(t)$ and $s(t)$, $I \int_{-\infty}^{t} V(t') - s(t') dt'$. In addition to this expected operation, the actual output signal is delayed by $\tau_{\mathrm{contr}}$ because of electronical response delay.

As written in the main text, the response of the AOD-controlled laser delays by $\tau_{\text{AOD}}$ compared to the input signal to the AOD $\varepsilon_{\text{AOD}}(t)$ (Fig. S1A). To evaluate $\tau_{\text{AOD}}$, we conducted the back-focal-plain laser interferometry (BFPI) using a probe particle adhered to the surface of the glass slide in the custom-made sample chamber (Fig. S2). The drive laser was focused on the particle, and its focus position $u_{\text{AOD}}$ was oscillated by supplying a sinusoidal signal $\varepsilon_{\text{AOD}}(t) = \hat{\varepsilon}_{\text{AOD}} \exp(-i\omega t)$ to the AOD. In this case, $u = 0$ and $u_{\text{AOD}} = u_{\text{d}}$ were always satisfied (see Fig.3B in the main text). The output voltage of the QPD for the drive laser was given by $V(t) = u_{\text{AOD}}(t)/C_{\text{d}} = \hat{u}_{\text{AOD}}/C_{\text{d}} \exp(-i\omega t)$. A lock-in amplifier was then used to measure the output of the QPD with $\varepsilon_{\text{AOD}}$ as the reference while $\hat{\varepsilon}_{\text{AOD}}$ was kept a real constant. As written in Ref. [1], the amplitude $|\hat{u}_{\text{AOD}}/C_{\text{d}}|$ did not remarkably change within the observed range of frequencies (0.01~$10^5$ Hz). On the other hand, the phase delay $\text{Arg}(\hat{u}_{\text{AOD}}/C_{\text{d}})$ was proportional to the applied frequency (Fig. S3A). Note that $\hat{u}_{\text{AOD}}$ is the complex function that depends on frequency as $\hat{u}_{\text{AOD}} = \hat{u}_{\text{AOD}}(\omega)$, and the phase of $\hat{u}_{\text{AOD}}$ reflects the time delay of QPD output relative to the AOD-input. When $u_{\text{AOD}}(t)$ is delayed from $\varepsilon_{\text{AOD}}(t)$ by $\tau_{\text{AOD}}$, *i.e.* $u_{\text{AOD}}(t) = C_{\text{AOD}} \varepsilon_{\text{AOD}}(t - \tau_{\text{AOD}}) = C_{\text{AOD}} \hat{\varepsilon}_{\text{AOD}} \exp(i\omega\tau_{\text{AOD}}) \exp(-i\omega t)$, the phase delay is obtained as $\text{Arg}(\hat{u}_{\text{AOD}}/C_{\text{d}}) = \omega\tau_{\text{AOD}}$. In Fig. S3A, the phase delay was fitted by $\text{Arg}(\hat{u}_{\text{AOD}}/C_{\text{d}}) = 2\pi f \tau_{\text{AOD}} = 4.16 \times 10^{-5} \times f$, which gave $\tau_{\text{AOD}} = 4.16/(2\pi) \times 10^{-5} = 6.62 \times 10^{-6}$ s.

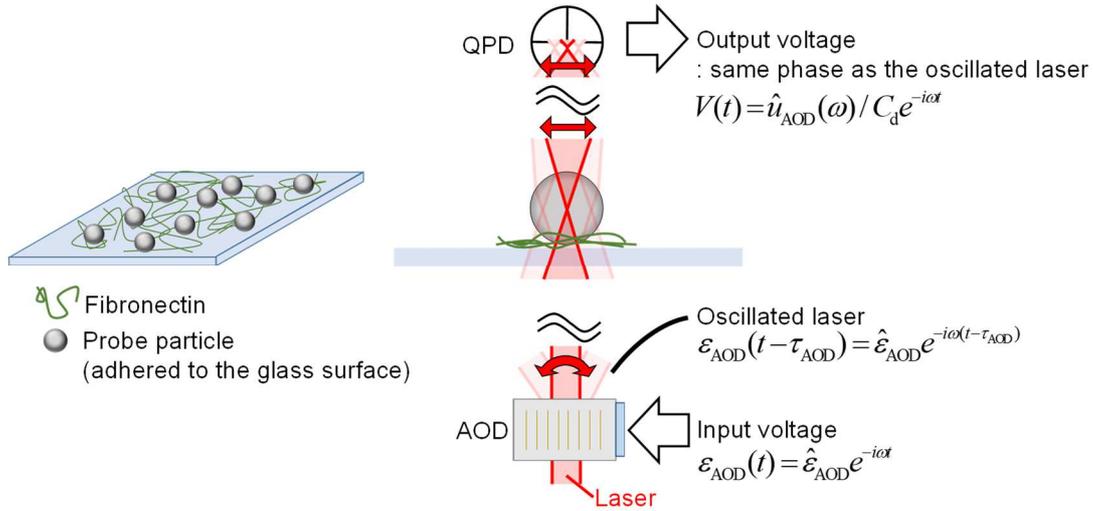

**Fig. S2** Setup for measurement of AOD-response delay $\tau_{\text{AOD}}$. Sinusoidal signal $\hat{\varepsilon}_{\text{AOD}} e^{-i\omega t}$ was input to AOD. The oscillating drive laser was diffracted by the colloidal particle (melamine, diameter = 1 μm) which was adhered to the glass surface of the sample chamber by fibronectin. Then, the QPD, which was set for the back-focal-plane laser interferometry, detected the diffraction. In the BFLI technique, the QPD output $V(t)$ is proportional to the laser oscillation $u_{\text{AOD}}(t)$ when the probe particle is fixed at the bottom of the slide glass. Then, the phase of QPD output signal is the same as that of oscillated laser $\text{Arg}[V(t)] = \text{Arg}[\varepsilon_{\text{AOD}}(t - \tau_{\text{AOD}})]$.

The PID feedback controller also generates a non-negligible delay; the output of the controller is delayed by $\tau_{\text{contr}}$ with respect to the signal fed to the "Measure" input (Fig. S1B). $\tau_{\text{contr}}$ was measured as follows. A sinusoidal signal $Ae^{-i\omega t}$ was fed to the Measure input of the PID controller while the set point was grounded. The output of the PID controller $Output(t)$ was measured by the lock-in amplifier with the measure input $Ae^{-i\omega t}$ as a reference. As shown in Fig. S3B, the phase delay $\text{Arg}(Output)$ was proportional to the input frequency $f$. When $\tau_{\text{contr}}$ is taken into consideration, the output of the PID controller is given by $Output(t) = IA \int_{-\infty}^{t-\tau_{\text{contr}}} \exp(-i\omega t')dt' = (IA/i\omega)\exp(i\omega\tau_{\text{contr}})\exp(-i\omega t)$. By fitting the phase delay in Fig. S3B with $\text{Arg}(Output) = \pi/2 + \omega\tau_{\text{contr}}$, $\tau_{\text{contr}} = 5.06/(2\pi) \times 10^{-6} = 8.1 \times 10^{-7}$ s was obtained.

Below, we investigate the effect of the time delays on MRs by systematically taking the delay times into account in our analysis. In Fig. S4, we show shear viscoelastic modulus $G(\omega)$ of 0.6 mg/mL F-actin gel (the same sample as that used in Fig. 3 in the main text) measured by conventional PMR, conventional AMR, and force-feedback AMR. A colloidal particle with $2a = 1$ μm diameter was used as a probe. All these time delays ($\tau_{\text{PID}}$, $\tau_{\text{contr}}$ and $\tau_{\text{AOD}}$) should be considered when force-feedback MR is analyzed whereas conventional AMR depends only on $\tau_{\text{AOD}}$. Obviously, conventional PMR is nothing related with these time delays. For conventional AMR, the phase delay in the measured response at an angular frequency $\omega$ is corrected by $\omega\tau_{\text{AOD}}$ [1]. As shown in Fig. S4, conventional AMR (closed symbols) and conventional PMR (curves) agreed well with each other as expected in a sample at equilibrium. As shown in eqn (18) in the main text, the displacement response of the probe with respect to the driving signal was affected by both $\tau_{\text{PID}}$ and $\tau \equiv \tau_{\text{AOD}} + \tau_{\text{contr}}$ when the force-feedback MR was conducted.

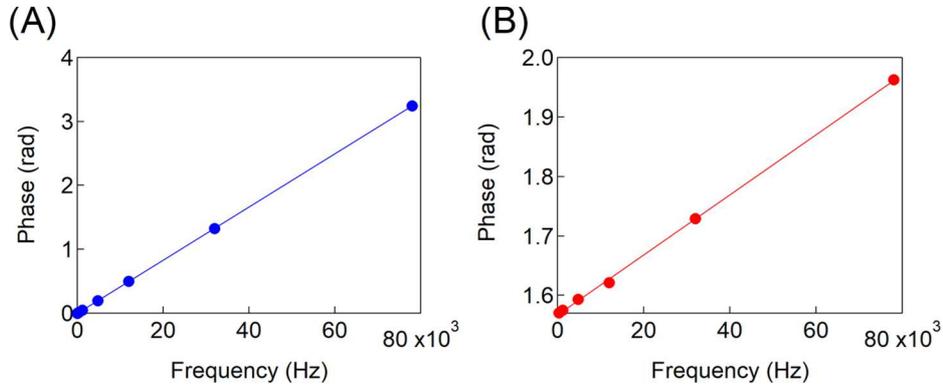

**Fig. S3** Phase delays observed in the response of (A) AOD and (B) PID feedback controller to the sinusoidal input signal. (A, B) Phase delays were proportional to the frequency as $\omega\tau_{\text{AOD}}$ and $\omega\tau_{\text{contr}}$, respectively. From the linear fit to the data (solid lines), we obtained $\tau_{\text{AOD}} = 6.6 \times 10^{-6}$ s and $\tau_{\text{contr}} = 8.1 \times 10^{-7}$ s.

Considering the experimental condition $\tau_{\text{contr}} < \tau_{\text{AOD}}(= 6.6 \times 10^{-6}$ s$) < \tau_{\text{PID}}(= 2.4 \times 10^{-5}$ s$)$, we first investigate the effect of time delays from the larger one in the force-feedback AMR experiment. Filled symbols in Fig. S4A were obtained by incorporating $\tau = 0$ in eqn (18) in the main text. Even if $\tau_{\text{AOD}}$ and $\tau_{\text{contr}}$ were neglected, $G(\omega)$ measured with force-feedback AMR mostly agreed with the results obtained by conventional AMR and PMR, except for the frequencies higher than 100 Hz (the yellow-colored region in Fig. S4A). Note that $G(\omega)$ at the highest frequency in force-feedback AMR was negative, and therefore it was not seen in the figure given by the logarithmic plot. This disagreement was decreased when $\tau = \tau_{\text{AOD}} = 6.6 \times 10^{-6}$ s and $\tau_{\text{PID}} = C_{\text{d}}/C_{\text{AOD}}I$ were considered in eqn (18) of the main text (filled symbols in Fig. S4B). Finally, the small delay time $\tau_{\text{contr}}$ was also taken into consideration as $\tau = \tau_{\text{AOD}} + \tau_{\text{contr}} = 7.4 \times 10^{-6}$ s. Then, the results of the force-feedback AMR agreed better with that of the conventional AMR (Fig. S4C). When $\tau_{\text{PID}}$ became comparable to $\tau$, the response delay $\tau$ of the feedback-controlled laser also affected force-feedback PMR experiments, which we will discuss in the supplementary 2.

The force-feedback AMR can operate when both $\omega \leq 1/\tau_{\text{PID}}$ and $\omega \leq 1/\tau$ are satisfied because the feedback-controlled laser does not run at the higher frequencies. In Fig. S4, the force-feedback AMR was performed up to 3 kHz because $\tau_{\text{PID}}$ was $2.4 \times 10^{-5}$ s. Even if $\omega \leq 1/\tau$ is held, our analysis indicates that the delay time $\tau = 7.4 \times 10^{-6}$ s is not negligible for the feedback operation of the AOD-controlled laser at large frequencies comparable to $1/2\pi\tau$ (yellow-colored region in Fig. S4). The influence of $\tau_{\text{contr}}$ on MR analysis becomes remarkable when the force feedback is run faster by setting $\tau_{\text{PID}} < \tau$. We discuss this situation in supplementary 2.

Although the conventional and force-feedback AMR were consistent at frequencies lower than 10 Hz, the force-feedback AMR provided more accurate and reasonable data. Note that $|G(\omega)|$ measured with conventional AMR systematically decreased as the frequency increased from 0.01 to 0.1 Hz. This frequency dependency is anomalous; it could occur only when the system (the probe particle and the surrounding medium) responds actively or autonomously to the applied external forces. Since $|G(\omega)|$ must monotonously increase with frequency in a sample at equilibrium, the observed decrease of $|G(\omega)|$ measured with conventional AMR was an artifact. It is necessary to correct the effect of the optical trapping on the probe movements when conventional AMR is performed. As discussed in prior studies and also in the main text of this study, the systematic error appears owing to the correction and it becomes greater at lower frequencies. Therefore, data taken at frequencies higher than 0.1 Hz was usually presented in previous studies when similar specimens were measured with the conventional AMR [1-3]. However, the artifacts disappeared in the force-feedback AMR because optical-trapping potential was effectively removed, as explained in the main text. Thus, we conclude that the force feedback technique revealed its expected performance of MR experiments at low frequencies.

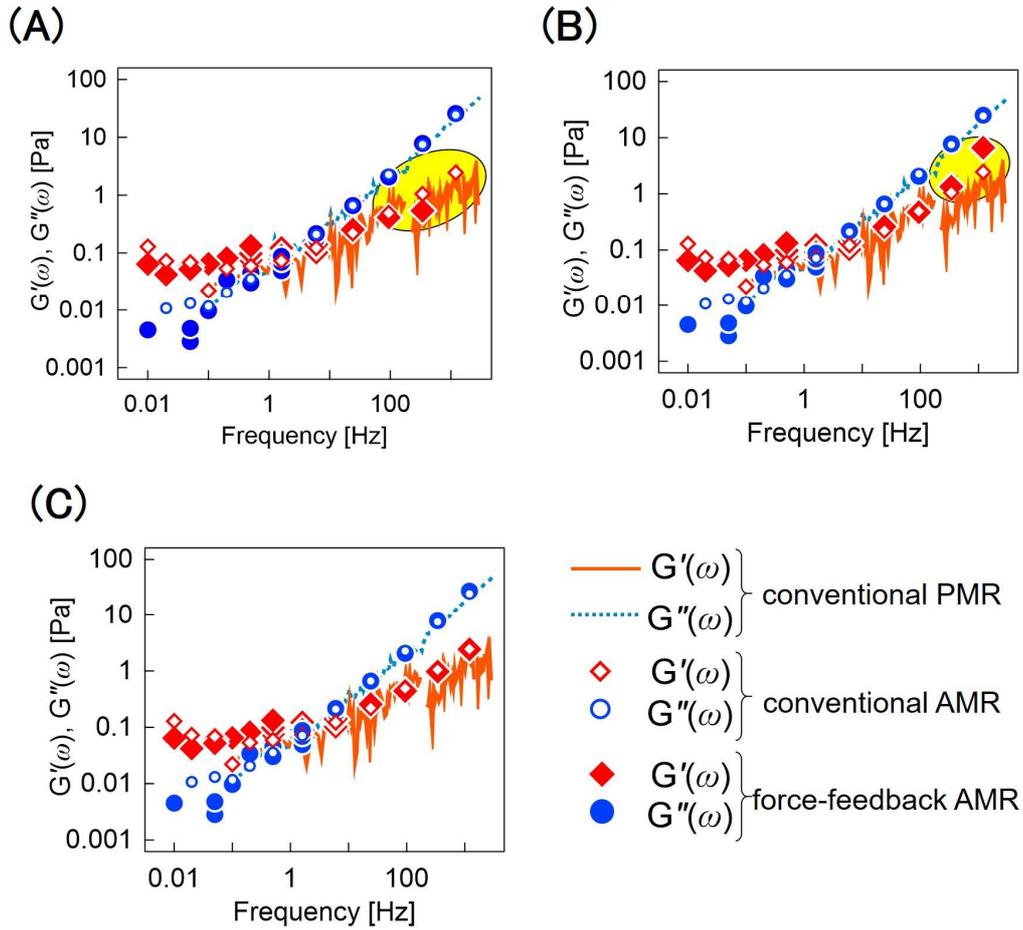

**Figure S4** Effect of time delays in the apparatus on the force-feedback MR. (A) Shear viscoelastic modulus $G(\omega)$ of 0.6 mg/mL entangled F-actin gel measured by conventional AMR (open symbols), conventional PMR (curves), and force-feedback AMR (filled symbols). Red and orange symbols: $G'(\omega)$. Blue and light blue symbols: $G''(\omega)$. The force-feedback AMR was analyzed by neglecting $\tau_{\mathrm{AOD}}$ and $\tau_{\mathrm{contr}}$. (B) The same as (A) except that $\tau_{\mathrm{AOD}}$ was taken into consideration when analyzing the force-feedback AMR (filled symbols). With this analysis, force-feedback AMR and conventional MR did not agree at frequencies higher than 100 Hz (see the yellow-colored region). (C) The same as Fig.3C in the main article, except for the addition of data measured by conventional AMR. Both $\tau_{\mathrm{AOD}}$ and $\tau_{\mathrm{contr}}$ were used to correct the force-feedback AMR (filled symbols).

## 2: Force-feedback PMR performed with small $\tau_{PID}$ comparable to $\tau_{AOD}$

In our feedback MR setup, the feedback-response time $\tau_{PID}$ can be adjusted by tuning the gain setting of the PID controller while $\tau_{AOD} \sim 10^{-5}\,\text{s}$ and $\tau_{contr} \sim 10^{-6}\,\text{s}$ were fixed. In section 1.2.1 in the main text, the force-feedback PMR was discussed by ignoring $\tau_{AOD}$ and $\tau_{contr}$ because $\tau_{PID}$ was set much larger than $\tau\,(=\tau_{AOD} + \tau_{contr})$ and was set much smaller than the characteristic response time for the trapped particle $\tau_c\,(\equiv \gamma_0 / k_p)$, i.e. $\tau \ll \tau_{PID} \ll \tau_c$. However, the performance of force-feedback PMR declines when the feedback is set faster by decreasing $\tau_{PID}$ down to the value comparable to $\tau$. In this section, we investigate how smaller $\tau_{PID}$ affects the force-feedback PMR experiments.

The force-feedback PMR was demonstrated using a $2a = 1\,\mu\text{m}$ latex particle in water. The AOD-controlled 1064 nm laser was used as probe. As $\tau_{PID}$ was decreased, the power spectral density (PSD) of $u_d(t)$ and $u_{AOD}(t)$ started to show a peak at $\sim 1/(2\pi\tau_{AOD})$ Hz (Fig. S5A). This peak grew when $\tau_{PID}$ was further decreased and approached $\tau_{AOD}$. Here, $u_{AOD}(t)$ was estimated as $u_{AOD}(t) = C_{AOD} \cdot \varepsilon(t - \tau_{AOD})$ based on the input signal to AOD [ $\varepsilon(t)$ ], whereas $u_d(t)$ was directly measured with BFPI. Note that the PSD of $u_{AOD}(t)$ can be estimated by measuring $\varepsilon(t)$ as $\left\langle |\tilde{u}_{AOD}(\omega)|^2 \right\rangle^{FF} = C_{AOD}^2 \cdot |\mathcal{F}[\varepsilon(t - \tau_{AOD})]|^2 = C_{AOD}^2 \cdot \left\langle |\tilde{\varepsilon}_{AOD}(\omega)|^2 \right\rangle^{FF}$ regardless of the time delay $\tau_{AOD}$. The force-feedback PMR experiment presented in the main text was conducted under the condition $\tau_{AOD} \ll \tau_{PID}$. In that case, the PSD of total probe displacement $u(t) = u_{AOD}(t) - u_d(t)$ was obtained by using eqn (10) in the main text which we rewrite here as:

$$\left\langle |\tilde{u}(\omega)|^2 \right\rangle^{FF} = \left\langle |\tilde{u}_d(\omega)|^2 \right\rangle^{FF} + \left\langle |\tilde{u}_{AOD}(\omega)|^2 \right\rangle^{FF}. \tag{S1}$$

In Fig. S5B, $\left\langle |\tilde{u}(\omega)|^2 \right\rangle^{FF}$ calculated using eqn (S1) is shown. When $\tau_{PID}$ is decreased, $\left\langle |\tilde{u}(\omega)|^2 \right\rangle^{FF}$ disagreed with the theoretical prediction made by the fluctuation-dissipation theorem (FDT)

$$\left\langle |\tilde{u}(\omega)|^2 \right\rangle = k_B T / 3\pi a \omega^2 \eta. \tag{S2}$$

Here, $a$ and $\eta$ are radius of probe particle and viscosity of water, respectively (eqn (11) in the main text). Note that eqn (S1) was valid only when $\tau_{AOD}$ was negligibly small ( $\tau_{AOD} \ll \tau_{PID}$ ) and the correlation between $u_{AOD}(t)$ and $u_d(t)$ was made as $\tilde{u}_d = -i\omega\tau_{PID}\tilde{u}_{AOD}$ by the feedback control (see eqn (9) in the main text). When $\tau_{AOD} \ll \tau_{PID}$ is not satisfied, the correlation underlying the eqn (S1) is altered by the delay time $\tau_{AOD}$. Below, we will provide more detailed analysis of the force-feedback PMR, and extend eqn (S1) to more general situation by taking $\tau_{AOD} \sim 10^{-5}\,\text{s}$ and $\tau_{contr} \sim 10^{-6}\,\text{s}$ into account.

First, we consider $\tau_{AOD}$ in the force-feedback PMR analysis. Since the displacement of the AOD-controlled laser is delayed from $\varepsilon_{AOD}$ by $\tau_{AOD}$, $u_{AOD}$ under the force-feedback PMR is given by

$$u_{AOD}(t) = C_{AOD}\,\varepsilon_{AOD}(t - \tau_{AOD}) = (1/\tau_{PID}) \int_{-\infty}^{t-\tau_{AOD}} u_d(t')\,dt'. \tag{S3}$$

This equation is also obtained by incorporating $s(t) = L\exp(-i\omega t) = 0$ and $\tau = \tau_{AOD}\,(\tau_{contr} = 0)$ in eqn (16)

in the main text. The Fourier transform of eqn (S3) is written as

$$\tilde{u}_{\text{AOD}}(\omega) = \int_{-\infty}^{\infty} C_{\text{AOD}} \varepsilon_{\text{AOD}}(t' - \tau_{\text{AOD}}) e^{i\omega t'} dt' = e^{i\omega\tau_{\text{AOD}}} \int_{-\infty}^{\infty} C_{\text{AOD}} \varepsilon_{\text{AOD}}(t'') e^{i\omega t''} dt''$$

$$= e^{i\omega\tau_{\text{AOD}}} C_{\text{AOD}} \tilde{\varepsilon}_{\text{AOD}}(\omega) = -\left(\frac{e^{i\omega\tau_{\text{AOD}}}}{i\omega\tau_{\text{PID}}}\right) \tilde{u}_{\text{d}}(\omega). \tag{S4}$$

Eqn (S4) indicates that the time lag $\tau_{\text{AOD}}$ does not affect the relation between the auto power spectrums $\left\langle |\tilde{u}_{\text{AOD}}(\omega)|^2 \right\rangle^{\text{FF}} = C_{\text{AOD}}^2 \cdot \left\langle |\tilde{\varepsilon}_{\text{AOD}}(\omega)|^2 \right\rangle^{\text{FF}}$, as discussed in the previous paragraph. By using eqn (S4), the PSD of the probe displacement $u(t)$ is derived as,

$$\left\langle |\tilde{u}(\omega)|^2 \right\rangle^{\text{FF}} = \left\langle |\tilde{u}_{\text{AOD}}(\omega) - \tilde{u}_{\text{d}}(\omega)|^2 \right\rangle^{\text{FF}} = \left\langle \left| \left\{ 1 + \left[ \frac{e^{i\omega\tau_{\text{AOD}}}}{i\omega\tau_{\text{PID}}} \right] \right\} \tilde{u}_{\text{d}}(\omega) \right|^2 \right\rangle^{\text{FF}}$$

$$= \left\langle \left| \left\{ 1 + \left( \frac{1}{\omega\tau_{\text{PID}}} \right)^2 + 2 \frac{\sin\omega\tau_{\text{AOD}}}{\omega\tau_{\text{PID}}} \right\} \tilde{u}_{\text{d}}(\omega) \right|^2 \right\rangle^{\text{FF}} \tag{S5}$$

$$= \left\langle |\tilde{u}_{\text{d}}(\omega)|^2 \right\rangle^{\text{FF}} + \left\langle |\tilde{u}_{\text{AOD}}(\omega)|^2 \right\rangle^{\text{FF}} + 2 \frac{\sin\omega\tau_{\text{AOD}}}{\omega\tau_{\text{PID}}} \left\langle |\tilde{u}_{\text{d}}(\omega)|^2 \right\rangle^{\text{FF}}.$$

We calculated $\left\langle |\tilde{u}(\omega)|^2 \right\rangle^{\text{FF}}$ under $\tau_{\text{PID}} = 1.8 \times 10^{-5}$ s by incorporating measured $\left\langle |\tilde{u}_{\text{d}}(\omega)|^2 \right\rangle^{\text{FF}}$ and $\left\langle |\tilde{u}_{\text{AOD}}(\omega)|^2 \right\rangle^{\text{FF}} = C_{\text{AOD}}^2 \cdot \left\langle |\tilde{\varepsilon}_{\text{AOD}}(\omega)|^2 \right\rangle^{\text{FF}}$ into eqn (S5) instead of eqn (S1). Thus obtained $\left\langle |\tilde{u}(\omega)|^2 \right\rangle^{\text{FF}}$ (red curve in Fig. S6) agreed well with the theoretical prediction (black curve in Fig. S6) made by the FDT eqn (S2) at all frequencies.

In force-feedback PMR, a probe particle is subjected to the friction force $-\int_{-\infty}^{t} \gamma(t - t') \dot{u}(t') dt'$,

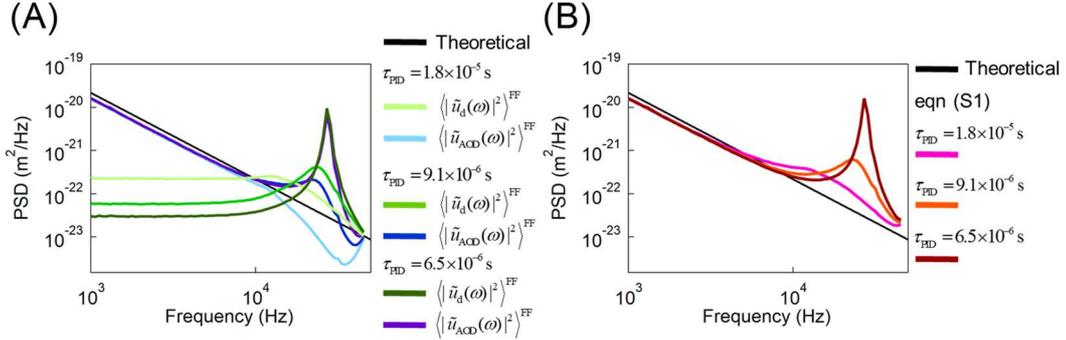

**Figure S5**: (A) Power spectral density (PSD) of $u_{\text{d}}(t)$ (green curves) and $u_{\text{AOD}}(t)$ (blue and purple curves) measured under the fast force feedback PMR conducted with small $\tau_{\text{PID}}$ in water. $\tau_{\text{PID}} : 1.8 \times 10^{-5}$ s, $9.1 \times 10^{-6}$ s and $6.5 \times 10^{-6}$ s. $\lambda = 1064$ nm laser was used as the probe laser and Latex particle with $2a = 1$ μm diameter was used as a probe. The delay time of AOD was $\tau_{\text{AOD}} = 7.51 \times 10^{-6}$ s. (B) PSD of probe displacement $\left\langle |\tilde{u}(\omega)|^2 \right\rangle^{\text{FF}}$ calculated by eqn (S1) with the data shown in (A). The solid black lines are the thermal fluctuations predicted by the fluctuation-dissipation theorem (Stokes-Einstein equation), $\left\langle |\tilde{u}(\omega)|^2 \right\rangle = k_{\text{B}} T / 3\pi a \omega^2 \eta$ [eqn (S2)], where $\eta$ is the viscosity of water at room temperature (1.01 mPa·s).

thermally fluctuation force $f(t)$, and the optical trapping force from probe laser $-k_{\mathrm{p}}u_{\mathrm{d}}(t)$. Then, Langevin equation for the probe particle is written as

$$\int_{-\infty}^{t} \gamma(t-t')\dot{u}(t')dt' = -k_{\mathrm{p}}u_{\mathrm{d}}(t) + f(t). \tag{S6}$$

Here, the friction function is given by $\gamma(t) = \gamma_0 \delta(t)$ using the friction coefficient $\gamma_0 = 6\pi a\eta$ of a probe particle in a Newtonian liquid (water). By using eqns (S4), (S5) and the FDT of the 2$^{\mathrm{nd}}$ kind $\left\langle \left| \tilde{f}(\omega) \right|^2 \right\rangle = 2k_{\mathrm{B}}T\gamma_0$, $\left\langle \left| \tilde{u}_{\mathrm{d}}(\omega) \right|^2 \right\rangle^{\mathrm{FF}}$ and $\left\langle \left| \tilde{u}_{\mathrm{AOD}}(\omega) \right|^2 \right\rangle^{\mathrm{FF}}$ are obtained,

$$\left\langle \left| \tilde{u}_{\mathrm{d}}(\omega) \right|^2 \right\rangle^{\mathrm{FF}} = \frac{2k_{\mathrm{B}}T\gamma_0}{\left( k_{\mathrm{p}} + \dfrac{\gamma_0 \cos(\omega\tau_{\mathrm{AOD}})}{\tau_{\mathrm{PID}}} \right)^2 + \omega^2\gamma_0^2 \left( 1 + \dfrac{\sin(\omega\tau_{\mathrm{AOD}})}{\omega\tau_{\mathrm{PID}}} \right)^2}, \tag{S7}$$

$$\left\langle \left| \tilde{u}_{\mathrm{AOD}}(\omega) \right|^2 \right\rangle^{\mathrm{FF}} = \frac{2k_{\mathrm{B}}T\gamma_0 \, / \, \omega^2\tau_{\mathrm{PID}}^2}{\left( k_{\mathrm{p}} + \dfrac{\gamma_0 \cos(\omega\tau_{\mathrm{AOD}})}{\tau_{\mathrm{PID}}} \right)^2 + \omega^2\gamma_0^2 \left( 1 + \dfrac{\sin(\omega\tau_{\mathrm{AOD}})}{\omega\tau_{\mathrm{PID}}} \right)^2}. \tag{S8}$$

When the force-feedback control is run sufficiently fast ($\tau_{\mathrm{PID}} \ll \gamma_0/k_{\mathrm{p}}$), $k_{\mathrm{p}}(= 6.72\times10^{-6}) \ll \gamma_0 \cos(\omega\tau_{\mathrm{AOD}})/\tau_{\mathrm{PID}} \approx 6\pi a\eta/\tau_{\mathrm{PID}} \sim 10^{-3}$ is satisfied at frequencies where $\omega\tau_{\mathrm{AOD}} \ll \pi/2$ is satisfied. Besides, $(k_{\mathrm{p}} + \gamma_0 \cos(\omega\tau_{\mathrm{AOD}})/\tau_{\mathrm{PID}})^2 \ll \omega^2\gamma_0^2 (1 + \sin(\omega\tau_{\mathrm{AOD}})/\omega\tau_{\mathrm{PID}})^2$ holds at $\omega\tau_{\mathrm{AOD}} \geq \pi/2$. Therefore, the effects of trap stiffness $k_{\mathrm{p}}$ on $\left\langle \left| \tilde{u}_{\mathrm{d}}(\omega) \right|^2 \right\rangle^{\mathrm{FF}}$ and $\left\langle \left| \tilde{u}_{\mathrm{AOD}}(\omega) \right|^2 \right\rangle^{\mathrm{FF}}$ are both negligible in the fast force-feedback PMR. Then, eqns (S7) and (S8) are simplified by neglecting the optical trapping force, $k_{\mathrm{p}}u_{\mathrm{d}}(t) \approx 0$ as,

$$\left\langle \left| \tilde{u}_{\mathrm{d}}(\omega) \right|^2 \right\rangle^{\mathrm{FF}} = \frac{\tau_{\mathrm{PID}}^2 k_{\mathrm{B}}T}{3\pi a\eta(1 + 2\omega\tau_{\mathrm{PID}}\sin\tau_{\mathrm{AOD}}\omega + \omega^2\tau_{\mathrm{PID}}^2)}, \tag{S9}$$

$$\left\langle \left| \tilde{u}_{\mathrm{AOD}}(\omega) \right|^2 \right\rangle^{\mathrm{FF}} = \frac{k_{\mathrm{B}}T}{3\pi a\eta(1 + 2\omega\tau_{\mathrm{PID}}\sin\tau_{\mathrm{AOD}}\omega + \omega^2\tau_{\mathrm{PID}}^2)\omega^2}, \tag{S10}$$

where Stokes law $\gamma_0 = 6\pi a\eta$ is used. Note that the equation consistent to the FDT [eqn (S2)] is obtained by incorporating eqn (S9) and (S10) into eqn (S5).

In Fig. S6, we show experimental results $\left\langle \left| \tilde{u}_{\mathrm{d}}(\omega) \right|^2 \right\rangle^{\mathrm{FF}}$, $\left\langle \left| \tilde{u}_{\mathrm{AOD}}(\omega) \right|^2 \right\rangle^{\mathrm{FF}}$ measured with $\tau_{\mathrm{PID}} = 1.8\times10^{-5}$ s $\gtrsim \tau_{\mathrm{AOD}}$ and corresponding theoretical predictions with eqns (S9) and (S10). Note that all parameters appearing in the right-hand side of eqns (S9) and (S10) were determined by independent experiments. Theoretical estimations of $\left\langle \left| \tilde{u}_{\mathrm{d}}(\omega) \right|^2 \right\rangle^{\mathrm{FF}}$ and $\left\langle \left| \tilde{u}_{\mathrm{AOD}}(\omega) \right|^2 \right\rangle^{\mathrm{FF}}$, which are shown by the broken curves in Fig. S6, agreed well with the experimental data (solid curves) without any adjustable parameters. Our theoretical analysis taking the delay time $\tau_{\mathrm{AOD}}$ into consideration seems to be valid at the condition $\tau_{\mathrm{PID}} = 1.8\times10^{-5}$ s.

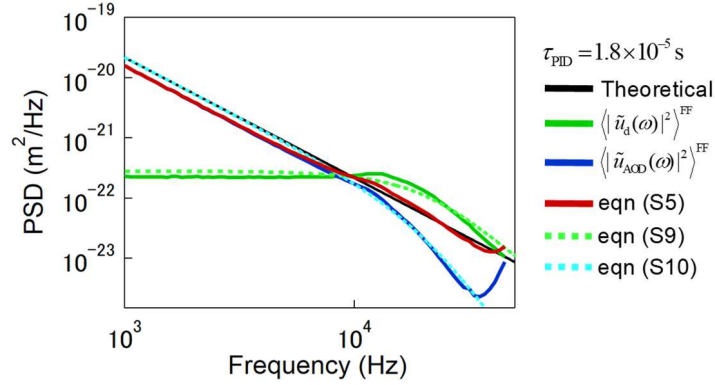

**Figure S6**: PSDs of $u(t)$, $u_{\mathrm{d}}(t)$, $u_{\mathrm{AOD}}(t)$ measured with force-feedback PMR ( $\tau_{\mathrm{AOD}} \lesssim \tau_{\mathrm{PID}} = 1.8 \times 10^{-5}$ s ). Solid curves are experimental results [ $\left\langle \left| \tilde{u}_{\mathrm{d}}(\omega) \right|^2 \right\rangle^{\mathrm{FF}}$ , $\left\langle \left| \tilde{u}_{\mathrm{AOD}}(\omega) \right|^2 \right\rangle^{\mathrm{FF}}$ ] which are the same as corresponding PSDs shown in Fig. S5A. $\left\langle \left| \tilde{u}_{\mathrm{d}}(\omega) \right|^2 \right\rangle^{\mathrm{FF}}$ (solid green curve) and $\left\langle \left| \tilde{u}_{\mathrm{AOD}}(\omega) \right|^2 \right\rangle^{\mathrm{FF}}$ (solid blue curve) agreed well with their theoretical predictions: eqn (S9) (broken light green curve) and eqn (S9) (broken light blue curve), respectively, indicating that $\tau_{\mathrm{contr}}$ is negligible under this condition. Consequently, PSD of the probe displacement $\left\langle \left| \tilde{u}(\omega) \right|^2 \right\rangle^{\mathrm{FF}}$ (solid red curve) also agreed well with the theoretical prediction eqn (S2) (solid black line).

Next, we discuss force-feedback PMR conducted with $\tau_{\mathrm{PID}} = 9.1 \times 10^{-6}$ s . For such a small $\tau_{\mathrm{PID}}$ , not only $\tau_{\mathrm{AOD}}$ but also the response-delay time $\tau_{\mathrm{contr}} (= 8.1 \times 10^{-7}$ s) of the PID controller may not be neglected, as we will see below. The time delay $\tau_{\mathrm{contr}}$ that emerges between input and output of PID controller (see supplementary 1) alters eqn (8) $C_{\mathrm{AOD}} \varepsilon_{\mathrm{AOD}}(t) = \frac{1}{\tau_{\mathrm{PID}}} \int_{-\infty}^{t} u_d(t') dt'$ in the main text as $C_{\mathrm{AOD}} \varepsilon_{\mathrm{AOD}}(t) = \frac{1}{\tau_{\mathrm{PID}}} \int_{-\infty}^{t-\tau_{\mathrm{contr}}} u_d(t') dt'$ . Then, the displacement of the laser $u_{\mathrm{AOD}}(t)$ is given by correcting eqn (S3),

$$ u_{\mathrm{AOD}}(t) = C_{\mathrm{AOD}} \, \varepsilon_{\mathrm{AOD}}(t - \tau_{\mathrm{AOD}}) = (1/\tau_{\mathrm{PID}}) \int_{-\infty}^{t-\tau_{\mathrm{AOD}}-\tau_{\mathrm{contr}}} u_d(t') dt' . . \tag{S11} $$

The Fourier transform of $u_{\mathrm{AOD}}(t)$ is

$$
\begin{aligned}
\tilde{u}_{\mathrm{AOD}}(\omega) &= \int_{-\infty}^{\infty} C_{\mathrm{AOD}} \varepsilon_{\mathrm{AOD}}(t - \tau_{\mathrm{AOD}}) e^{i\omega t} dt = \frac{1}{\tau_{\mathrm{PID}}} \int_{-\infty}^{\infty} e^{i\omega t} dt \int_{-\infty}^{t-\tau_{\mathrm{AOD}}-\tau_{\mathrm{contr}}} u_{\mathrm{d}}(t') dt' \\
&= -\frac{1}{i\omega\tau_{\mathrm{PID}}} \int_{-\infty}^{\infty} u_{\mathrm{d}}(t - \tau_{\mathrm{AOD}} - \tau_{\mathrm{contr}}) e^{i\omega t} dt = -\frac{1}{i\omega\tau_{\mathrm{PID}}} \int_{-\infty}^{\infty} u_{\mathrm{d}}(t'') e^{i\omega(t''+\tau_{\mathrm{AOD}}+\tau_{\mathrm{contr}})} dt'' \\
&= -\frac{e^{i\omega(\tau_{\mathrm{AOD}}+\tau_{\mathrm{contr}})}}{i\omega\tau_{\mathrm{PID}}} \int_{-\infty}^{\infty} u_{\mathrm{d}}(t'') e^{i\omega t''} dt'' = -\frac{e^{i\omega(\tau_{\mathrm{AOD}}+\tau_{\mathrm{contr}})}}{i\omega\tau_{\mathrm{PID}}} \tilde{u}_{\mathrm{d}}(\omega).
\end{aligned}
\tag{S12}
$$

Therefore, the exact formula for the PSD of probe displacement $\left\langle \left| \tilde{u}(\omega) \right|^2 \right\rangle^{\mathrm{FF}}$ that takes both $\tau_{\mathrm{AOD}}$ and $\tau_{\mathrm{contr}}$ into account is

$$\left\langle \left| \tilde{u}(\omega) \right|^2 \right\rangle^{\mathrm{FF}} = \left\langle \left| \tilde{u}_{\mathrm{d}}(\omega) \right|^2 \right\rangle^{\mathrm{FF}} + \left\langle \left| \tilde{u}_{\mathrm{AOD}}(\omega) \right|^2 \right\rangle^{\mathrm{FF}} + 2 \frac{\sin \omega(\tau_{\mathrm{AOD}} + \tau_{\mathrm{contr}})}{\omega \tau_{\mathrm{PID}}} \left\langle \left| \tilde{u}_{\mathrm{d}}(\omega) \right|^2 \right\rangle^{\mathrm{FF}}. \qquad (S13)$$

Accordingly, exact expressions for $\left\langle \left| \tilde{u}_{\mathrm{d}}(\omega) \right|^2 \right\rangle^{\mathrm{FF}}$ and $\left\langle \left| \tilde{u}_{\mathrm{AOD}}(\omega) \right|^2 \right\rangle^{\mathrm{FF}}$ are obtained by correcting eqns (S9) and (S10) in the same way:

$$\left\langle \left| \tilde{u}_{\mathrm{d}}(\omega) \right|^2 \right\rangle^{\mathrm{FF}} = \frac{\tau_{\mathrm{PID}}^2 k_{\mathrm{B}} T}{3 \pi a \eta (1 + 2\omega \tau_{\mathrm{PID}} \sin(\tau_{\mathrm{AOD}} + \tau_{\mathrm{contr}}) \omega + \omega^2 \tau_{\mathrm{PID}}^2)}, \qquad (S14)$$

$$\left\langle \left| \tilde{u}_{\mathrm{AOD}}(\omega) \right|^2 \right\rangle^{\mathrm{FF}} = \frac{k_{\mathrm{B}} T}{3 \pi a \eta (1 + 2\omega \tau_{\mathrm{PID}} \sin(\tau_{\mathrm{AOD}} + \tau_{\mathrm{contr}}) \omega + \omega^2 \tau_{\mathrm{PID}}^2) \omega^2}. \qquad (S15)$$

Note that eqns (S13)-(S15) are obtained by exchanging $\tau_{\mathrm{AOD}}$ in the corresponding eqns (S5), (S9) and (S10) to the total delay time $\tau = \tau_{\mathrm{AOD}} + \tau_{\mathrm{contr}}$.

Fig. S7A shows PSDs of probe displacement $\left\langle \left| \tilde{u}(\omega) \right|^2 \right\rangle^{\mathrm{FF}}$ calculated using eqns (S5) and (S13) using the same data [ $\left\langle \left| \tilde{u}_{\mathrm{d}}(\omega) \right|^2 \right\rangle^{\mathrm{FF}}$ and $\left\langle \left| \tilde{u}_{\mathrm{AOD}}(\omega) \right|^2 \right\rangle^{\mathrm{FF}}$ ] measured at $\tau_{\mathrm{PID}} = 9.1 \times 10^{-6}$ s. PSDs shown in Fig. S7B were normalized to $\omega^2 \left\langle \left| \tilde{u}(\omega) \right|^2 \right\rangle^{\mathrm{FF}}$ to clarify the effect of $\tau_{\mathrm{contr}}$. The PSD obtained by considering both $\tau_{\mathrm{AOD}}$ and $\tau_{\mathrm{contr}}$ (orange squares) agreed better with the theoretical prediction (black curve) based on the FDT [eqn (S2)] than the PSD estimated by neglecting $\tau_{\mathrm{contr}}$ (red triangles). We therefore conclude that the response of our force-feedback systems was properly understood even if the characteristic response time $\tau_{\text{PID}}$ is as small as the delay time of our instruments $\tau = \tau_{\text{AOD}} + \tau_{\text{contr}}$, and the dynamics of the probe particle were successfully recovered if we consider all the characteristic times: $\tau_{\text{PID}}$, $\tau_{\text{contr}}$ and $\tau_{\text{AOD}}$.

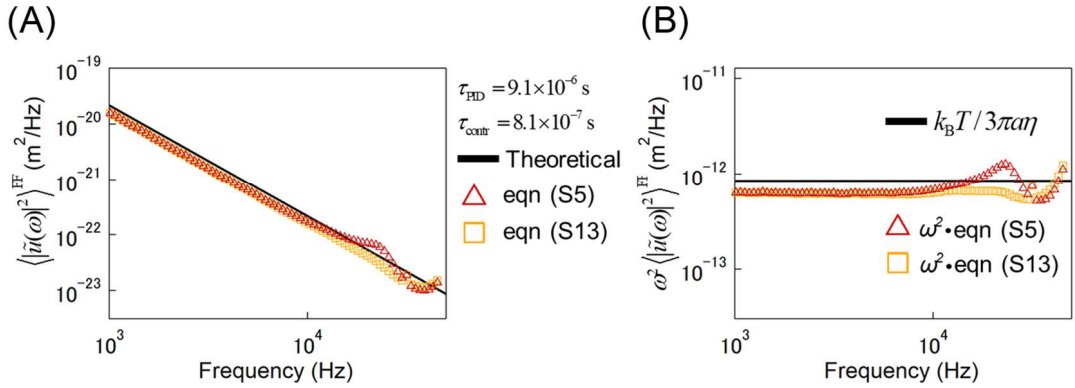

**Figure S7**: PSD of a probe particle under the force feedback with $\tau_{\mathrm{AOD}} \sim \tau_{\mathrm{PID}} = 9.1 \times 10^{-6}$ s. Experimental results were analyzed with and without $\tau_{\mathrm{contr}}$. (A) PSDs of probe displacement $\left\langle \left| \tilde{u}(\omega) \right|^2 \right\rangle^{\mathrm{FF}}$ were obtained using the corresponding data shown in Fig. S5A. Red triangles were calculated by eqn (S5) that neglected $\tau_{\mathrm{contr}}$. Orange squares were calculated by eqn (S13) considering all time delays including $\tau_{\mathrm{contr}} = 8.1 \times 10^{-7}$ s (see supplementary 1). (B) PSDs normalized as $\omega^2 \left\langle \left| \tilde{u}(\omega) \right|^2 \right\rangle^{\mathrm{FF}}$. Red triangles and orange squares were obtained using eqn (S5) and eqn (S13), respectively. Solid black lines in (A) and (B) are the theoretical prediction $\omega^2 \left\langle \left| \tilde{u}(\omega) \right|^2 \right\rangle = k_{\mathrm{B}} T / 3 \pi a \eta$. Under this condition ( $\tau_{\mathrm{AOD}} \sim \tau_{\mathrm{PID}}$ ), the analysis including all time delays showed better agreement with the theoretical prediction except a slight underestimation which is constant with frequency.

Finally, the force feedback was operated even faster (*i.e.* $\tau_{\text{PID}} = 6.5 \times 10^{-6}$ s $\sim \tau_{\text{AOD}}$), and the PSD of the probe particle was evaluated by incorporating $\left\langle \left| \tilde{u}_{\text{d}}(\omega) \right|^2 \right\rangle^{\text{FF}}$ and $\left\langle \left| \tilde{u}_{\text{AOD}}(\omega) \right|^2 \right\rangle^{\text{FF}}$ (Fig. S8A) into eqn (S13) (pink squares in Fig. S8B). $\left\langle \left| \tilde{u}(\omega) \right|^2 \right\rangle^{\text{FF}}$ could not be accurately evaluated even if $\tau_{\text{contr}}$ and $\tau_{\text{AOD}}$ were considered. Under such an extremely fast feedback, slight errors in the estimation of $\tau_{\text{AOD}}$ and $\tau_{\text{contr}}$ could prevent the estimation of PSD for $\omega \geq 1/\tau_{\text{AOD}}$. We then fitted the theoretical model [eqns (S14) and (S15)] to data [ $\left\langle \left| \tilde{u}_{\text{d}}(\omega) \right|^2 \right\rangle^{\text{FF}}$ and $\left\langle \left| \tilde{u}_{\text{AOD}}(\omega) \right|^2 \right\rangle^{\text{FF}}$ ] by choosing $\tau$ as an adjustable parameter. Results with $\tau = 8.64 \times 10^{-6}$ s are shown by solid curves in Fig. S8A. Although the model seems to fit well with our data with reasonable value of $\tau$, $\left\langle \left| \tilde{u}(\omega) \right|^2 \right\rangle^{\text{FF}}$ evaluated by using eqn (S13) with the same $\tau = 8.64 \times 10^{-6}$ s disagreed with the theoretical prediction by the FDT $\left\langle \left| \tilde{u}(\omega) \right|^2 \right\rangle = k_{\text{B}}T/3\pi a \omega^2 \eta$ [black curve, eqn (S2)] at high frequencies as shown by red triangles in Fig. S8B.

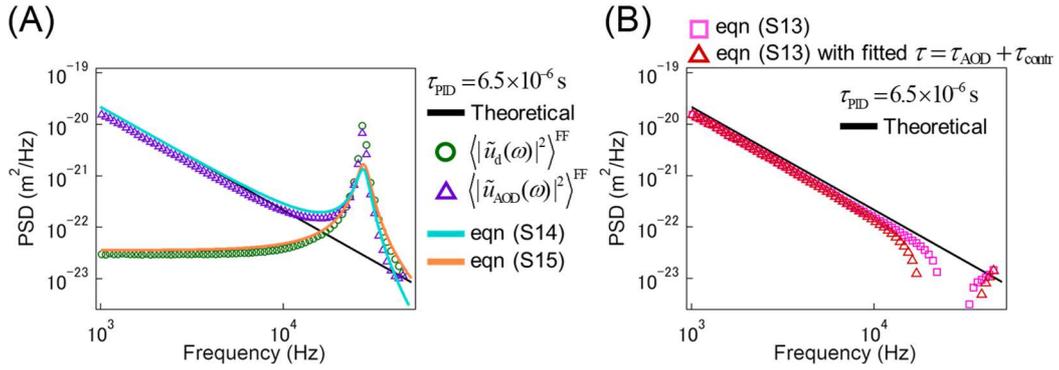

**Figure S8**: Fit of the theoretical model to force-feedback PMR data measured at $\tau_{\text{PID}} = 6.5 \times 10^{-6}$ s. (A) PSD of $u_{\text{d}}(t)$ and $u_{\text{AOD}}(t)$ measured with $\tau_{\text{PID}} = 6.5 \times 10^{-6}$ s (green circles and purple triangles, the same as those in Fig. S5A) were fitted with eqns (S14) and (S15) (light blue and orange curves). $\tau$ $(= \tau_{\text{AOD}} + \tau_{\text{contr}})$ was chosen as a fitting parameter and adjusted to be $8.64 \times 10^{-6}$ s. (B) PSD of probe displacement $\left\langle \left| \tilde{u}(\omega) \right|^2 \right\rangle^{\text{FF}}$ calculated by eqn (S13). Pink squares: PSD calculated with experimentally obtained parameters $\tau_{\text{AOD}} (= 7.5 \times 10^{-6}$ s$)$ and $\tau_{\text{contr}} (= 8.1 \times 10^{-7}$ s$)$. Red triangles: PSD calculated with fitted $\tau = 8.64 \times 10^{-6}$ s.

In the force-feedback PMR experiments given in the main text, we obtained the probe displacement from the measured $\varepsilon_{\text{AOD}}(t)$ and $u_{\text{d}}(t)$ as $u(t) = u_{\text{AOD}}(t) - u_{\text{d}}(t) = C_{\text{AOD}} \cdot \varepsilon_{\text{AOD}}(t) - u_{\text{d}}(t)$, and then its PSD was calculated. This was possible because the condition $\tau_{\text{AOD}} \ll \tau_{\text{PID}}$ was satisfied. When $\tau_{\text{AOD}}$ is not negligible compared to $\tau_{\text{PID}}$, $\left\langle \left| \tilde{u}_{\text{d}}(\omega) \right|^2 \right\rangle^{\text{FF}}$ and $\left\langle \left| \tilde{u}_{\text{AOD}}(\omega) \right|^2 \right\rangle^{\text{FF}} = C_{\text{AOD}}^2 \cdot \left\langle \left| \tilde{\varepsilon}_{\text{AOD}}(\omega) \right|^2 \right\rangle^{\text{FF}}$ were calculated first, and then eqn (S13) was used to obtain $\left\langle \left| \tilde{u}(\omega) \right|^2 \right\rangle^{\text{FF}}$ as explained above. Instead, it is also possible to obtain

$u(t)$ as

$$u(t) = u_{AOD}(t) - u_d(t) = C_{AOD} \cdot \varepsilon_{AOD}(t - \tau_{AOD}) - u_d(t) , \qquad (S16)$$

by shifting the $\varepsilon_{AOD}(t)$ by the time delay $\tau_{AOD}$ as $u_{AOD}(t) = C_{AOD} \cdot \varepsilon_{AOD}(t - \tau_{AOD})$ . Its PSD is then obtained as $\left\langle |\tilde{u}(\omega)|^2 \right\rangle^{FF} = \left\langle |\mathcal{F}[u_{AOD}(t) - u_d(t)]|^2 \right\rangle^{FF}$ . This method seems to be of advantage because it does not depend on $\tau_{PID}$ and $\tau_{contr}$ . However, $\left\langle |\tilde{u}(\omega)|^2 \right\rangle^{FF}$ thus obtained also did not conform to the theoretical prediction $\left\langle |\tilde{u}(\omega)|^2 \right\rangle = k_B T / 3\pi a \omega^2 \eta$ (Fig. S9). In this study, $u_d(t)$ and $\varepsilon_{AOD}(t)$ were sampled at the interval of $\sim 10^{-5}$ s, which was larger than the time delay of AOD ( $\tau_{AOD} = 7.5 \times 10^{-6}$ s). Therefore, because of the lack of the time resolution, it was not possible to accurately estimate the probe displacement $u(t)$ by using eqn (S16).

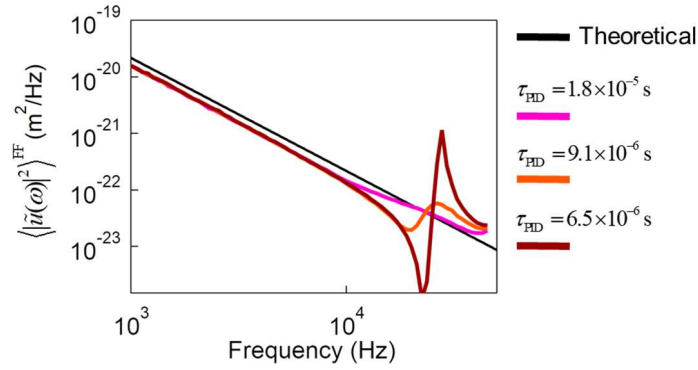

**Figure S9**: PSDs of probe particle $\left\langle |\tilde{u}(\omega)|^2 \right\rangle^{FF}$ in water measured by force-feedback PMR. $\left\langle |\tilde{u}(\omega)|^2 \right\rangle^{FF}$ were analyzed with the consideration of $\tau_{AOD}$ (= $7.5 \times 10^{-6}$ s) by using the relation $\left\langle |\tilde{u}(\omega)|^2 \right\rangle^{FF} = \left\langle |\mathcal{F}[u_{AOD}(t) - u_d(t)]|^2 \right\rangle^{FF} = \left\langle |\mathcal{F}[C_{AOD} \cdot \varepsilon_{AOD}(t - \tau_{AOD}) - u_d(t)]|^2 \right\rangle^{FF}$ at $\tau_{PID} = 1.8 \times 10^{-5}$ s, $9.1 \times 10^{-6}$ and $6.5 \times 10^{-6}$ s (pink, orange, and red curves, respectively). Experimental data used in the calculation $u_d(t)$, $C_{AOD} \cdot \varepsilon_{AOD}(t)$ are same as those of Fig. S5A. Black curve shows the theoretical prediction [eqn (S2)].

In general, it becomes harder to evaluate PSD of the probe particle $\left\langle |\tilde{u}(\omega)|^2 \right\rangle^{FF}$ accurately when force-feedback PMR was operated with $\tau_{PID}$ smaller than $\tau$ . When $\tau_{PID}$ was decreased, the PSD of $\varepsilon_{AOD}(t)$ [ $\left\langle |\tilde{\varepsilon}_{AOD}(\omega)|^2 \right\rangle^{FF}$ ] did not decay at $\omega \geq 1/\tau$ , but instead, it showed a large peak at $\omega \sim 1/\tau$ , meaning that the laser is required to vigorously oscillate around the frequency. However, at this range of frequencies, the spacing of the diffraction grating in AOD (see Fig. S1A) will not be a constant within the volume where the laser deflects, as discussed in a prior study [1]. Consequently, the laser deflection not only delays from the control voltage $\varepsilon_{AOD}(t)$ , but also may nonlinearly alter the amplitude; *i.e*, $C_{AOD}$ in the relation $u_{AOD}(t) = C_{AOD} \cdot \varepsilon(t - \tau_{AOD})$ may no longer be a constant. The nonlinearity in the response of AOD may be amplified when the control voltage $\varepsilon_{AOD}(t)$ has large oscillation amplitude, as indicated by the peak in $\left\langle |\tilde{\varepsilon}_{AOD}(\omega)|^2 \right\rangle^{FF}$ . This artifact coming from the nonlinearity of the AOD is hard to analyze, and therefore it determines the limit of the fast feedback control for MR experiments.

## 3: Slow force-feedback PMR at which slow probe particle's movement is suppressed by optical-trapping force.

Until here in this supplementary, we have investigated the fast force-feedback MR where $\tau_{\text{PID}}$ satisfies the condition $\tau = \tau_{\text{AOD}} + \tau_{\text{contr}} \sim \tau_{\text{PID}} \ll \tau_{\text{c}} = \gamma_0 / k_{\text{p}}$. In this section, we will consider the PMR conducted under the slower force feedback. First, the force-feedback PMR was performed with $\tau_{\text{PID}} = 1.8 \times 10^{-4}\,\text{s}$ while other experimental conditions were kept the same as those in the previous section. Because $\tau_{\text{PID}}$ is much larger than $\tau_{\text{AOD}}$, the PSD of the probe displacement under force feedback [ $\left\langle |\tilde{u}(\omega)|^2 \right\rangle^{\text{FF}}$ ] can be evaluated by using eqn (S1). Results are shown by red triangles in Fig. S10A, and they agree well with the theoretical prediction $\left\langle |\tilde{u}(\omega)|^2 \right\rangle = k_{\text{B}}T / 3\pi a \omega^2 \eta$ [eqn (S2)] shown by the black line. Under the condition for this experiment ( $\tau \ll \tau_{\text{PID}} \ll \tau_{\text{c}}$ ), both the effect of optical trapping force ( $k_p u_d$ ) and the time delays ( $\tau_{\text{AOD}}$ and $\tau_{\text{contr}}$ ) are negligible. In such case, $\left\langle |\tilde{u}(\omega)|^2 \right\rangle^{\text{FF}}$ estimated by eqn (S1) is equivalent to eqn (S2), which is the optimal situation for conducting the force-feedback PMR.

Next, the force feedback PMR was performed by further increasing $\tau_{\text{PID}}$ up to $\tau_{\text{PID}} = 1.8 \times 10^{-3}\,\text{s}$. As shown in Fig. S10B, $\left\langle |\tilde{u}(\omega)|^2 \right\rangle^{\text{FF}}$ calculated by eqn (S1) (red triangles) systematically deviates from the theoretical prediction [eqn (S2), black line] at low frequencies. Under the slow force feedback ( $\tau_{\text{PID}} > \tau_{\text{c}} \gg \tau_{\text{AOD}}$ ), eqn (S1) accurately estimates the PSD [ $\left\langle |\tilde{u}(\omega)|^2 \right\rangle^{\text{FF}}$ ] since $\tau_{\text{AOD}}$ and $\tau_{\text{contr}}$ are negligible as is the case in Fig. S10A. However, it differs from the thermal fluctuation of a "free" probe particle which is not subjected to the optical trapping.

Here, we quantitatively discuss the fluctuation of the probe particle under the slow feedback to understand the observed inconsistency between $\left\langle |\tilde{u}(\omega)|^2 \right\rangle$ and $\left\langle |\tilde{u}(\omega)|^2 \right\rangle^{\text{FF}}$. We again start with the Langevin equation for the probe particle,

$$\int_{-\infty}^{t} \gamma(t-t')\dot{u}(t')dt' = -k_p u_d(t) + f(t), \tag{S17}$$

where the fluctuation-dissipation theorem of the 2$^{\text{nd}}$ kind

$$\left\langle |\tilde{f}(\omega)|^2 \right\rangle = 2k_{\text{B}}T \, \text{Re}\left[ \tilde{\gamma}(\omega) \right] = \frac{2k_{\text{B}}T \alpha''(\omega)}{\omega |\alpha(\omega)|^2} \tag{S18}$$

should hold for the thermal fluctuating force. For a probe particle dispersed in a viscoelastic continuum, the memory function and the intrinsic response function are related as $\tilde{\gamma}(\omega) = -1/i\omega\alpha(\omega)$. PSD of the probe particle $\left\langle |\tilde{u}(\omega)|^2 \right\rangle^{\text{FF}}$ is calculated by incorporating eqns (S18) and (S4) into the Fourier transformation of eqn (S17) as

$$\left\langle |\tilde{u}(\omega)|^2 \right\rangle^{\text{FF}} = \frac{2k_{\text{B}}T \alpha''(\omega)}{\omega |1 - \beta\alpha(\omega)|^2} = \frac{1}{|1 - \beta\alpha(\omega)|^2} \left\langle |\tilde{u}(\omega)|^2 \right\rangle. \tag{S19}$$

Here, $\beta$ is defined as $\beta \equiv k_p / (1 + 1/i\omega\tau_{\text{PID}})$ and $\tau_{\text{AOD}} = \tau_{\text{contr}} = 0$ is used. $\left\langle |\tilde{u}(\omega)|^2 \right\rangle = 2k_{\text{B}}T \alpha''(\omega)/\omega$ is the FDT of the 1$^{\text{st}}$ kind, representing the thermal PSD for a free probe particle dispersed in a viscoelastic continuum. In the case of purely viscous liquid, incorporating $\alpha(\omega) = -1/6\pi i\omega\eta a$ leads to eqn (S2).

$\left\langle |\tilde{u}(\omega)|^2 \right\rangle$ calculated by using eqn (S19) agreed well with the theoretical curve

$\left\langle \left| \bar{u}(\omega) \right|^2 \right\rangle = k_\mathrm{B} T / 3\pi a \omega^2 \eta$ [eqn (S2)] (open pink squares in Fig. S10B), including the low-frequency region where $\left\langle \left| \bar{u}(\omega) \right|^2 \right\rangle^{\mathrm{FF}}$ deviates from eqn (S2). As it is written in the main text, force feedback effectively eliminates the resilient force from the optical-trapping potential when the feedback control is fast (meaning $\tau_\mathrm{PID} \ll \tau_\mathrm{c}$). When the feedback is slow ( $\tau_\mathrm{PID} > \tau_\mathrm{c}$ ), the probe motion is suppressed by the optical-trapping potential. In the case of a probe particle trapped by a fixed laser, its PSD typically shows a plateau for $\omega < 1/\tau_\mathrm{c}$ [1]. Similarly, $\left\langle \left| \bar{u}(\omega) \right|^2 \right\rangle^{\mathrm{FF}}$ also starts to show the similar plateau for $\omega < 1/\tau_\mathrm{c}$, deviating from $\left\langle \left| \bar{u}(\omega) \right|^2 \right\rangle$ as seen in Fig. S10B. However, the plateau disappeared for $\omega < 1/\tau_\mathrm{PID}$ and recovered the diffusion-like dependency $\left\langle \left| \bar{u}(\omega) \right|^2 \right\rangle^{\mathrm{FF}} \propto 1/\omega^2$ because the feedback-controlled laser [ $u_\mathrm{AOD}(t)$ ] follows the thermal fluctuation of the probe in that range of frequencies.

In this section and prior sections, we have discussed the effect of time delays and optical trapping on force feedback MR, and shown how to correct those artifacts. For actual experiments, it is important to set the experimental condition to satisfy $\tau_\mathrm{AOD} \ll \tau_\mathrm{PID} \ll \tau_\mathrm{c}$ to avoid such tedious corrections. However, it is not always possible to choose appropriate $\tau_\mathrm{PID}$ prior to experiments. $\tau_\mathrm{c}$ is not a constant for general viscoelastic medium, and is not known before measurements. The range for $\tau_\mathrm{PID}$ satisfying $\tau_\mathrm{AOD} \ll \tau_\mathrm{PID} \ll \tau_\mathrm{c}$ might not be so broad especially when the strong laser is used. Only by thoroughly understanding the probe dynamics under force feedback, as we had done in this study, it is possible to identify the artifacts and correct them by using *e.g.* eqn (S19).

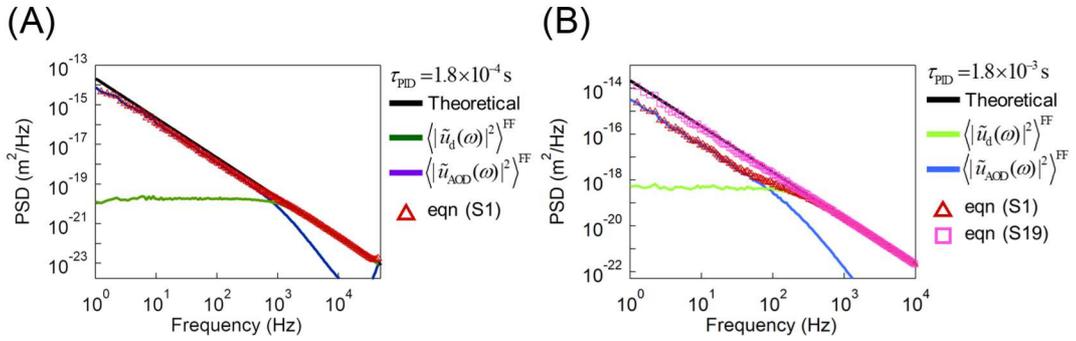

**Figure S10**: PSDs measured with force-feedback PMR in water, which was performed under the same experimental condition as that in Fig.S5 except for greater $\tau_\mathrm{PID}$ . (A) PSDs with the condition $\tau_\mathrm{AOD} \ll \tau_\mathrm{PID} \ll \tau_\mathrm{c} = \gamma_0 / k_\mathrm{p}$ . $\tau_\mathrm{PID} = 1.8 \times 10^{-4}\,\mathrm{s}$ and $\tau_\mathrm{c} = 1.4 \times 10^{-3}\,\mathrm{s}$ . Since $\tau_\mathrm{AOD}$ was negligible, $\left\langle \left| \bar{u}(\omega) \right|^2 \right\rangle^{\mathrm{FF}}$ estimated with eqn (S1) agreed well with the theoretical estimation for the thermal fluctuation $\left\langle \left| \bar{u}(\omega) \right|^2 \right\rangle = k_\mathrm{B} T / 3\pi a \omega^2 \eta$ [solid black line, eqn (S2)]. (B) PSDs with the condition $\tau_\mathrm{PID} \sim \tau_\mathrm{c}$ . $\tau_\mathrm{PID} = 1.8 \times 10^{-3}\,\mathrm{s}$ . $\left\langle \left| \bar{u}(\omega) \right|^2 \right\rangle^{\mathrm{FF}}$ calculated with eqn (S1) (red triangles) deviated from theoretical estimation $\left\langle \left| \bar{u}(\omega) \right|^2 \right\rangle = k_\mathrm{B} T / 3\pi a \omega^2 \eta$ (solid black line) at low frequencies because the probe fluctuation was suppressed by optical trapping. The corrected PSD of probe displacement (pink squares) calculated from eqn (S19) showed good agreement with the theoretical estimation (solid black line).

# 4: The sensitivity and offset error in BFPI signal depend on the laser focus position

In the case of force-feedback MR using a single AOD-controlled laser, the laser could move away from the optical axis during an experiment. Then, the sensitivity $1/C_d$ and offset-error voltage in QPD output $V_0$ may vary. Therefore, we measured $1/C_d$ and $V_0$ as a function of the laser deflection by AOD and estimated the errors that may appear during feedback-MR experiments.

Melamine particles with a diameter of 1 μm were trapped in water, and the focus position of the laser was oscillated by AOD as $u_{AOD}(t) = Ae^{-i\omega t} + u_L$, with $A = 71.5$ nm and $\omega/2\pi = 10$ kHz. $u_L$ is the average distance of the laser focus from the optical axis (Fig. S11A). During the experiment, the probe particle was trapped at the position $u_L$ because the oscillation of the laser was much faster than the response time of the probe ($1/\tau_c = k_p / \gamma_0 \simeq 130$ Hz). Therefore, the separation $u_d$ between the laser focus and the probe center was oscillated sinusoidally as $u_d = Ae^{-i\omega t}$. By measuring the QPD output voltage $V(t; u_L) = A/C_d(u_L)e^{-i\omega t} + V_0(u_L)$ with the lock-in amplifier, the sensitivity $1/C_d$ [V/m] was obtained as the ratio between the oscillation amplitudes of $u_{AOD}(t)$ and $V(t; u_L)$. The offset of the QPD output $V_0(u_L)$ was obtained by taking the time average of the QPD output, $V_0(u_L) = \langle V(t; u_L) \rangle$.

In Fig. S11 (B), the red circles and open blue triangles represent the sensitivity ($1/C_d$) and the offset ($V_0$), respectively. When $|u_L|$ was increased, the sensitivity $1/C_d$ tended to decrease, and the offset error $V_0$ arose. The laser deflection by the AOD is certified up to 45 mrad by the manufacturer, which corresponds to $-10$ μm $\leq u_L \leq 25$ μm in our setup. Within the range, $1/C_d$ varied more than 10% and $|V_0|$ exceeded 1 V at large $|u_L|$. When calibrated, $|V_0| \sim 1$ V corresponds to more than 100 nm. BFPI accurately measures the probe displacement when the probe laser is fixed. When the probe laser was moved more than 10 μm, it is common to have alteration of $1/C_d$ and $V_0$ similar to those observed here. The accuracy of particle tracking during the force feedback thus depends on the movement of the probe laser that follows the probe fluctuation. The force-feedback PMR in the main text was accurately performed because the probe dispersed in a highly viscous sample in which thermal Brownian motion was small. When a probe particle fluctuates vigorously, dual-feedback technique should be used to keep the laser movement within a certain limit that should be determined by the accuracy required for the measurement. Because the feedback-controlled stage tracks the slow and large movement of the probe particle, the AOD-controlled laser is kept close to the optical axis. Therefore, the dual-feedback technique is necessary to perform BFPI accurately when a probe particle vigorously fluctuates or drifts in samples driven out of equilibrium.

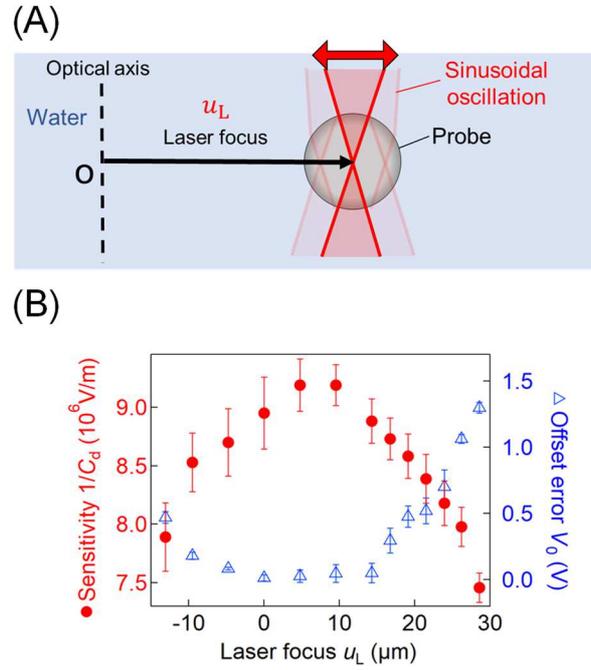

**Figure S11** (A) A schematic illustration of the experiment to evaluate the dependence of sensitivity $1/C_d$ and offset error $V_0$ on the focus position ($u_L$). The melamine particle with a diameter of 1 μm was dispersed in water, and trapped by the oscillating laser ($\lambda = 1064$ nm). By analyzing the QPD output, $1/C_d$ and $V_0$ were obtained as a function of $u_L$. (B) Dependence of $1/C_d$ (red circles) and $V_0$ (open blue triangles) on $u_L$. $1/C_d$ decreased and $V_0$ increased when the position of the laser focus was deviated from the center of the optical axis. At the largest $u_L$ in this measurement ($u_L = 28.6$ μm), $V_0 \simeq 1.3$ V corresponded to 0.2 μm when calibrated. The position with the highest sensitivity was shifted from the center of the optical axis probably because the center of the operating range of the AOD-controlled laser did not match the optical axis.

# 5: Scatter plot of step size $\Delta x$ and waiting time $t_w$ in the forced hopping of a probe partcile

As discussed in the main text, the dynamics of probe particles dispersed in the medium with heterogeneous microenvironments are described by introducing the density of states for the potential depth $E$ that microenvironments provide (Bouchaud's trap model). The power-law distributed $P_{wtd}(t_w)$ which is characteristic of glassy dynamics emerges when the density of states $\rho(E)$ is largely distributed. It is then reasonable to examine whether the work done for uncaging the microenvironments ($F\Delta x$) in a crosslinked F-actin gel may positively correlate with the waiting time $t_w$ taken before the step. The step size $\Delta x$ and the waiting time $t_w$ were obtained from the trajectory of the tracer beads subjected to a constant optical-trapping force ($F = 3.4$ pN), using the step detection algorithm mentioned in the main text. Pair of the data (the step size and the waiting time before the step) are shown in Fig. S12 by the scatter plot. No correlation between these quantities was observed while the step sizes and waiting times are both largely distributed.

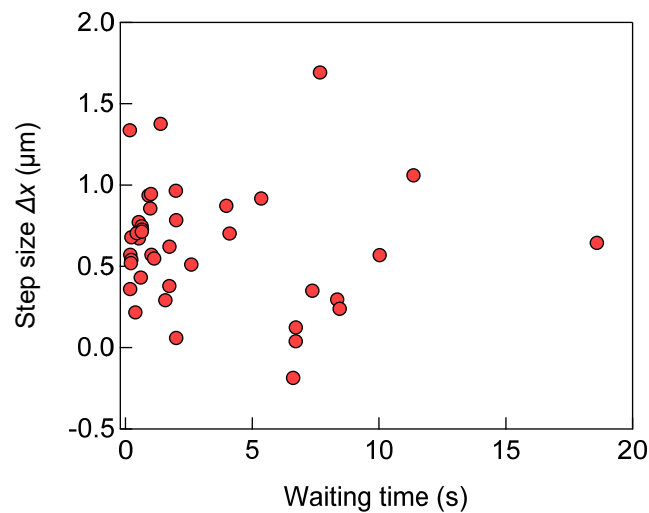

**Figure S12**: Scatter plot of step size $\Delta x$ and waiting time $t_w$ taken before the step event. The tracer beads in a crosslinked F-actin gel were under constant forcing $F = 4.3$pN (Fig 5C in the main text).

References

1.      D. Mizuno, D. A. Head, F. C. MacKintosh and C. F. Schmidt, *Macromolecules*, 2008, **41**, 7194-7202.

2.      D. Mizuno, C. Tardin and C. F. Schmidt, *Soft Matter*, 2020, **16**, 9369-9382.

3.      S. Jabbari-Farouji, M. Atakhorrami, D. Mizuno, E. Eiser, G. H. Wegdam, F. C. Mackintosh, D. Bonn and C. F. Schmidt, *Phys Rev E Stat Nonlin Soft Matter Phys*, 2008, **78**, 061402.